\documentclass[fleqn,usenatbib]{mnras}

\usepackage{comment}
\usepackage{newtxtext,newtxmath}

\usepackage[T1]{fontenc}
\usepackage{ae,aecompl}
\usepackage{booktabs}

\usepackage{graphicx}	
\usepackage{amsmath}	
\usepackage{amssymb}	
\usepackage{caption}
\usepackage{subfig}

\title[UNIT project: theoretical nonlinear models]{UNIT project: Universe $N$-body simulations for the Investigation of Theoretical models from galaxy surveys} 

\author[Chuang et al.]{
\parbox{\textwidth}{
Chia-Hsun Chuang$^{1}$\thanks{E-mail: chuangch@stanford.edu},
Gustavo Yepes$^{2,3}$\thanks{E-mail: gustavo.yepes@uam.es},
Francisco-Shu Kitaura$^{4,5}$,
Marcos Pellejero-Ibanez$^{4,5}$,
Sergio~Rodr\'iguez-Torres$^{2,3}$,
Yu Feng$^{6}$,
R. Benton Metcalf$^{7,8}$,
Risa H. Wechsler$^{1,9,10}$
Cheng Zhao$^{11,12}$,
Chun-Hao To$^{1,9,10}$,
Shadab Alam$^{13}$,
Arka Banerjee$^{1,9,10}$,
Joseph DeRose$^{1,9,10}$,
Carlo Giocoli$^{8,14,7,15}$,
Alexander Knebe$^{2,3,16}$,
Guillermo Reyes$^{2,3}$
}
\vspace*{4pt} \\
$^{1}$Kavli Institute for Particle Astrophysics and Cosmology, Stanford University, 452 Lomita Mall, Stanford, CA 94305, USA\\
$^{2}$Departamento de F\'isica Te\'{o}rica, M\'{o}dulo 8, Facultad de Ciencias, Universidad Aut\'{o}noma de Madrid, 28049 Madrid, Spain\\
$^{3}$CIAFF, Facultad de Ciencias, Universidad Aut\'{o}noma de Madrid, 28049 Madrid, Spain\\
$^{4}$Instituto de Astrof\'{\i}sica de Canarias (IAC), C/V\'{\i}a L\'actea, s/n, E-38200, La Laguna, Tenerife, Spain \\
$^{5}$Departamento Astrof\'{\i}sica, Universidad de La Laguna (ULL), E-38206 La Laguna, Tenerife, Spain \\
$^{6}$Berkeley Center for Cosmological Physics, Department of Physics, University of California Berkeley, Berkeley CA, 94720, USA\\
$^{7}$Dipartimento di Fisica e Astronomia, Alma Mater Studiorum Universit\`{a} di Bologna, via Gobetti 93/2, I-40129 Bologna, Italy\\
$^{8}$INAF - Osservatorio di Astrofisica e Scienza dello Spazio di Bologna,  via Gobetti 93/3, I-40129 Bologna, Italy \\
$^{9}$Department of Physics, Stanford University, 382 Via Pueblo Mall, Stanford, CA 94305, USA\\
$^{10}$SLAC National Accelerator Laboratory, 2575 Sand Hill Road, Menlo Park, CA 94025, USA\\
$^{11}$Laboratoire d'Astrophysique, Ecole Polytechnique F\'ed\'erale de Lausanne (EPFL), Observatoire de Sauverny, CH-1290 Versoix, Switzerland\\
$^{12}$National Astronomical Observatories, Chinese Academy of Science, Beijing 100012, P.R.China\\
$^{13}$Institute for Astronomy, University of Edinburgh, Royal Observatory, Blackford Hill, Edinburgh, EH9 3HJ , UK\\
$^{14}$Dipartimento di Fisica e Scienze della Terra, Universit\`a degli Studi di Ferrara, via Saragat 1, I-44122 Ferrara, Italy\\
$^{15}$INFN - Sezione di Bologna, viale Berti Pichat 6/2, I-40127
  Bologna, Italy\\
$^{16}$ICRAR, University of Western Australia, 35 Stirling Highway, Crawley, Western Australia 6009, Australia
}

\date{Accepted XXX. Received YYY; in original form ZZZ}

\pubyear{2018}

\begin{document}
\label{firstpage}
\pagerange{\pageref{firstpage}--\pageref{lastpage}}
\maketitle
%
%
\begin{abstract}
We present the UNIT $N$-body cosmological simulations project, designed to provide precise predictions for nonlinear statistics of the galaxy distribution.  We focus on characterizing statistics relevant to emission line and luminous red galaxies in the current and upcoming generation of galaxy surveys. We use a suite of precise particle mesh simulations (FastPM) as well as with full $N$-body calculations with a mass resolution of $\sim 1.2\times10^9\,h^{-1}$M$_{\odot}$ to investigate the recently suggested technique of Angulo \& Pontzen 2016 to suppress the variance of cosmological simulations  We study redshift space distortions, cosmic voids, higher order statistics from $z=2$ down to $z=0$.  We find that both two- and three-point statistics are unbiased. Over the scales of interest for baryon acoustic oscillations and redshift-space distortions, we find that the variance is greatly reduced in the two-point statistics and in the cross correlation between halos and cosmic voids, but is not reduced significantly for the three-point statistics.  We demonstrate that the accuracy of the two-point correlation function for a galaxy survey with effective volume of 20 ($h^{-1}$Gpc)$^3$ is improved by about a factor of 40, indicating that two pairs of simulations with a volume of 1 ($h^{-1}$Gpc)$^3$ lead to the equivalent variance of $\sim$150 such simulations.  The $N$-body simulations presented here thus provide an effective survey volume of about seven times the effective survey volume of DESI or Euclid. The data from this project, including dark matter fields, halo catalogues, and their clustering statistics, are publicly available: \url{http://www.unitsims.org}.
\end{abstract}





\section{Introduction}

The large-scale structure measured in galaxy surveys represents one of the most powerful probes of present day cosmology and of the nature of dark matter and dark energy in the Universe. To this end, a considerable observational effort is being put forward to  map the three-dimensional galaxy distribution in the Universe at unprecedented scales with large photometric and spectroscopic surveys that will measure the positions of tens to hundreds of millions of galaxies. 
The current largest photometric and spectroscopic surveys are the Dark Energy Survey\footnote{\url{http://www.darkenergysurvey.org}} (DES) and the Extended Baryon Oscillation Spectroscopic Survey \footnote{\url{http://www.sdss.org/sdss-surveys/eboss/}} (eBOSS) respectively.
The total volume of the Universe mapped with galaxy surveys is dramatically increasing, with several large upcoming ground- and space-based experiments being planned, including as
4MOST\footnote{\url{http://www.4most.eu/}} (4-metre Multi-Object Spectroscopic Telescope, \citealt{deJong:2012nj}), DESI\footnote{\url{http://desi.lbl.gov/}} (Dark Energy Spectroscopic Instrument, \citealt{Schlegel:2011zz,Levi:2013gra}),
HETDEX\footnote{\url{http://hetdex.org}} (Hobby-Eberly Telescope Dark Energy Experiment, \citealt{Hill:2008mv}),
J-PAS\footnote{\url{http://j-pas.org}} (Javalambre Physics of accelerating universe Astrophysical Survey, \citealt{Benitez:2014ibt}),
PFS\footnote{\url{https://pfs.ipmu.jp}}(Subaru Prime Focus Spectrograph, \citealt{2014PASJ...66R...1T}),
LSST\footnote{\url{http://www.lsst.org/lsst/}} (Large Synoptic Survey Telescope, \citealt{Abell:2009aa}), 
Euclid\footnote{\url{http://www.euclid-ec.org}} \citep{Laureijs:2011gra}, 
and WFIRST\footnote{\url{http://wfirst.gsfc.nasa.gov}} (Wide-Field Infrared Survey Telescope, \citealt{Spergel:2013tha}).  

In order to extract cosmological constraints from these surveys as well as the data allow, the systematic errors associated with theoretical models that for example characterize the galaxy power spectrum or correlation function as a function of cosmological model must be well below the statistical uncertainties caused by cosmic variance and shot noise. Some pioneering analytical models have been developed to compute the theoretical expected  correlation function. To date, these models have limited accuracy, as they rely on analytical gravity models \citep[e.g.][]{1970A&A.....5...84Z}, simplified biasing descriptions, and approximate redshift-space distortion models \citep[see][and references therein]{2015MNRAS.450.3822W}. These models have not achieved the accuracy possible with a numerical computation using a full gravity solver. To meet the goals of current galaxy surveys, simulations with much larger effective volumes than those probed by the surveys {\em and} with enough mass resolution to resolve the dark matter halos hosting the typical galaxies  detected in those surveys are required.  Yet, the computational resources needed to accomplish this task are at the edge of the current (petaflop) computational power.
A single simulation with the required halo mass resolution ($\sim 4\times10^{11}\,h^{-1}$M$_{\odot}$, \citealt{2017MNRAS.469.2913C}) that covers the whole volume sampled by Euclid ($\sim$ 70 Gpc$^3$) would demand an enormous number of particles (more than $16,000^3$ in a 4 Gpc$h^{-1}$ box).  The largest $N$-body simulations performed so far, e.g., 
MillenniumXXL  \citep{Angulo:2012ep},
MICE \citep{Fosalba:2013wxa},
MultiDark \citep{Klypin:2014kpa},
Dark Sky \citep{Skillman:2014qca}, 
OuterRim \citep{Habib:2014uxa}, and
FLAGSHIP \citep{Potter:2016ttn},
are still well below this particle number despite their computational expense. 

In this project, we explore an alternative way of reaching the same level of required accuracy using far fewer computational resources. 
Recently, \cite{Angulo:2016hjd} proposed a new method to dramatically reduce cosmic  variance arising from the sparse sampling of large-scale wave modes in cosmological simulations. The method uses pairs of simulations \citep{Pontzen:2015eoh} with initial Fourier-mode amplitudes that are fixed  to  the ensemble-averaged power spectrum and initial modes that are  exactly out of phase (one of the pair of simulations has opposite phases with respect to its companion). Using this  methodology, one can potentially obtain a result that is statistically equivalent to the mean of many independent simulations from a single pair of simulations. 

To date, the method has been tested only on the dark matter distribution at high redshifts ($z=1$) obtained with particle mesh gravity solvers, with a low-resolution particle mass of 1.7 $\times10^{12}\,h^{-1}$ M$_\odot$.
Given that fixed initial conditions cease to be formally a Gaussian field with a fixed power spectrum, there has been concern about potential biases this approach could introduce in the clustering statistics, although \cite{Angulo:2016hjd} gave analytical arguments why the biases should be negligible.
\cite{Villaescusa-Navarro:2018bpd} further tested this method using hydrodynamical simulations.  These simulations were done with small volumes (20 $h^{-1}$ Mpc on a side) or with low resolution (particle mass of 6.6 $\times10^{11}\,h^{-1}$ M$_\odot$) compared to the requirements of current and upcoming surveys. 

There is a need to directly test the usefulness and applicability of this approach to key large-scale structure analyses, including baryon acoustic oscillations and redshift-space distortions, that are expected with upcoming large surveys.  That is the goal of the present work.  Here, we extend these studies to the statistics, redshift range, galaxy samples, resolution, and volume required by surveys such as DESI and Euclid.
We use volumes of ($1 h^{-1}$Gpc)$^3$ in our studies.  Such volumes have been claimed to be large enough to account for large-scale mode coupling \citep{Klypin:2018ydv}, although this likely needs to be further investigated; missing modes may need to be accounted for in a post-processing step (Chuang et al in prep).

We have designed the simulations in the present work to focus on the key cosmology samples of upcoming surveys, which require robust modeling that encompasses the
expected halo masses for emission line galaxies (ELGs) ($\sim 10^{11} h^{-1}$M$_{\odot}$, \citealt{Gonzalez-Perez:2017mvf}) and H$\alpha$ galaxies ($\sim 4\times10^{11}\,h^{-1}$M$_{\odot}$, \citealt{2017MNRAS.469.2913C}). The simulation boxes are 1 $h^{-1}$Gpc on a side, with $4096^3$ particles and a mass resolution of $1.2 \times 10^9\,h^{-1}$M$_{\odot}$. We are thus able to safely resolve all halos with masses larger than $1.2\times10^{11}\,h^{-1}$M$_{\odot}$, using 100 particles per halo.

We demonstrate that the resulting errors in the statistical  correlation function measurements using the suppressed variance method are equivalent to having more than 7 times the effective volume sampled by DESI or Euclid galaxies ($\sim$20 ($h^{-1}$Gpc)$^3)$.  We generate halo catalogs and merger trees using the publicly available {\sc ROCKSTAR} halo finder \citep{Behroozi13}, together with density and velocity fields on a mesh for later construction of light-cone distributions of galaxies and weak lensing maps.
In future work, we will use these simulations to produce
thousands of catalogs, including mock galaxies with various techniques. 
The corresponding data will be made publicly available through databases and web portals for the general use of the astrophysical community\footnote{see \url{http://www.unitsims.org}}. 

This paper is organized as follows. First we present our study of the potential systematic biases from Suppressed Variance Methods (hereafter SVM; Section \ref{sec:fastpm}). In Section \ref{sec:gadget}, we present our suppressed variance simulation products including a clustering analysis and a robust assessment of the improvement. We summarize and conclude in Section \ref{sec:conclusion}. Throughout this work we use the 
following cosmological parameters: $\Omega_{\rm m}=0.3089$, $h\equiv H_0/100=0.6774$, $n_s=0.9667$ and $\sigma_8=0.8147$ (see Table 4 in \citealt{2016A&A...594A..13P}).

\section{Assesment of potential systematic biases in SVM}
\label{sec:fastpm}
We begin by studying the potential systematic biases and the improvement introduced by the suppressed variance method.
To this end, we want to generate a large total volume and number of simulations that permit us to estimate the uncertainties of the measurements and to quantify the improvements in the uncertainties on  different clustering measurements.

To create large simulated volumes, we rely on accelerated particle--mesh solvers, which have been recently shown to produce accurate halo populations compared to full $N$-body calculations, when enhanced with various techniques (see the COLA code \citealt{Tassev:2013pn} or the FastPM code \citealt{Feng:2016yqz}). 
 
\subsection{Setup}

We use the C implementation of the FastPM software, which employs a pencil domain-decomposition Poisson solver and a Fourier-space four-point differential kernel to compute the force. The time integration scheme is modified from a vanilla leap-frog scheme to account for the acceleration of velocity during a step and thus to correctly track the linear growth of large-scale modes regardless of the number of time steps.

\begin{table*}
\begin{tabular}{@{}|c|c|c|c|c|c|c|c|@{}}
\toprule
{\bf simulation}  & {\bf amplitude} & {\bf phases} & {\bf box side}        & {\bf number of } &{\bf particle} & {\bf force} & {\bf number of }                 \\ {\bf code}&&&{\bf length}&{\bf particles}&$M$ [$h^{-1}\,M_{\odot}$]& {\bf resolution}&{\bf boxes} \\ \midrule \midrule
Gadget G      & fixed   &regular      & 1 $h^{-1}$Gpc   & $4096^3$         &$1.2\times10^9$    &6 $h^{-1}$kpc &  2                                 \\ \midrule
Gadget $\overline{\rm G}$      & fixed   & inverse-phase of G    & 1 $h^{-1}$Gpc   & $4096^3$           &$1.2\times10^9$ & 6 $h^{-1}$kpc  & 2  
     \\ \midrule
FastPM A      & non-fixed & regular-reference LR & 1 $h^{-1}$Gpc   & $1024^3$   &$7.68\times10^{10}$ &  1.46  $h^{-1}$Mpc       & 100                         \\ \midrule
FastPM $\overline{\rm A}$      & non-fixed & inverse-phase of A & 1 $h^{-1}$Gpc   & $1024^3$          &$7.68\times10^{10}$  &1.46  $h^{-1}$Mpc &  100              \\ \midrule
FastPM B      & fixed   &  regular     & 1 $h^{-1}$Gpc   & $1024^3$ & $7.68\times10^{10}$      &  1.46  $h^{-1}$Mpc    & 100                              \\ \midrule
FastPM $\overline{\rm B}$      & fixed & inverse-phase of B    & 1 $h^{-1}$Gpc   & $1024^3$ & $7.68\times10^{10}$          &1.46  $h^{-1}$Mpc &  100              \\ \midrule
FastPM C      & non-fixed & regular-reference HR   & 250 $h^{-1}$Mpc & $1024^3$    &$1.2\times10^9$       & 0.36  $h^{-1}$Mpc  & 100                     \\ \midrule
FastPM $\overline{\rm C}$      & non-fixed  & inverse-phase of C & 250 $h^{-1}$Mpc & $1024^3$           &$1.2\times10^9$  &0.36  $h^{-1}$Mpc &  100             \\ \midrule
FastPM D      & fixed  & regular    & 250 $h^{-1}$Mpc & $1024^3$         &$1.2\times10^9$    &0.36  $h^{-1}$Mpc &  100                                 \\ \midrule
FastPM $\overline{\rm D}$      & fixed    & inverse-phase of D    & 250 $h^{-1}$Mpc & $1024^3$           &$1.2\times10^9$  &0.36  $h^{-1}$Mpc &  100           \\ 
\bottomrule
\end{tabular}
\caption{Overview of the set of simulations performed for this study and their corresponding parameter settings, including 800 FastPM and 2 pairs of Gadget simulations. LR and HR refer to low and high resolution, respectively.}\label{tab:sims}
\end{table*}

This permits us to efficiently perform 800 paired simulations to benchmark the method. Half of them (400) have  low resolution ($1024^3$ particles) but large volume (1 $h^{-1}$ Gpc side), and the other half (400) have enough resolution for the purpose of this study ($1024^3$ particles, see Section \ref{sec:gadget}) but smaller volume (250 $h^{-1}$ Mpc side). In each case we run half of the paired simulations with normal Gaussian random field initial conditions and half with fixed amplitudes, yielding 100 pairs of simulations for each case (see Table \ref{tab:sims}).

The simulations are started at $a\equiv 1/(1+z)=0.01$ ($z=99$), and evolved to $a=1$ ($z=0$) with 100 timesteps. We save snapshots of the simulations at $z=2$, $z=1$ and $z=0$. These regular simulation sets in high and low resolution define our reference simulations (see Table \ref{tab:sims}), from which we compute the standard mean summary statistics.

When computing halos with the Friends-of-Friends halo finder in {\tt nbodykit}\citep{Hand:2017pqn}, we choose a minimum of 20 dark matter particles per halo and a linking length of $0.2\, L_{\rm box}/N_{\rm c}$. Here $L_{\rm box}$ refers to the size of one side of the simulation box and $N_{\rm c}$ to the number of cells along one axis
used in the mesh computation, which was taken to correspond to the number of dark matter particles.  

\label{sec:fastpm_results}

\begin{figure*} \subfloat[PK-particles-FastPM\label{fig:1G_DM_PK_real_z1}]{\includegraphics[width=0.33\textwidth]{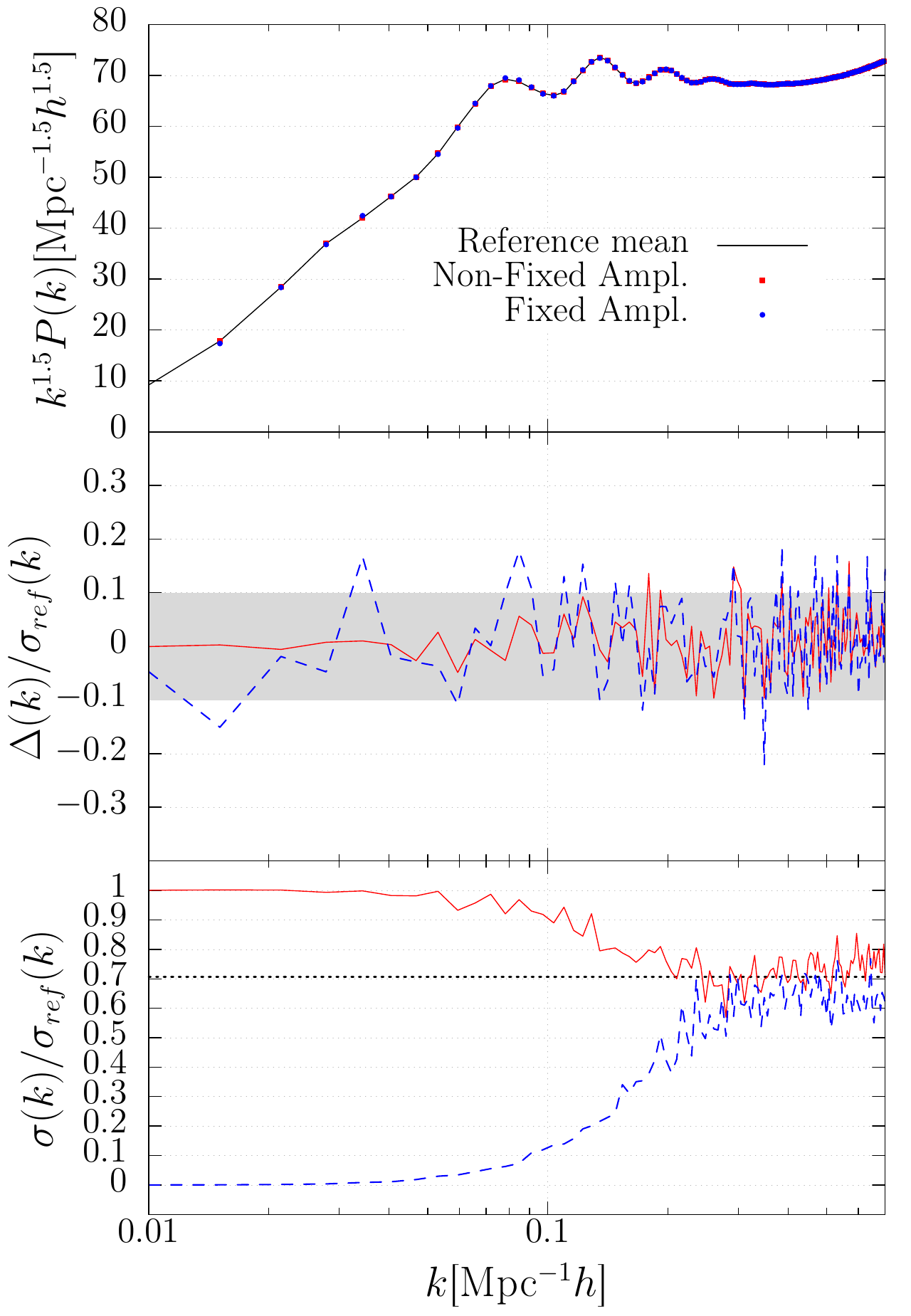}} \subfloat[CF-particles-FastPM\label{fig:1G_DM_CF_real_z1}]{\includegraphics[width=0.33\textwidth]{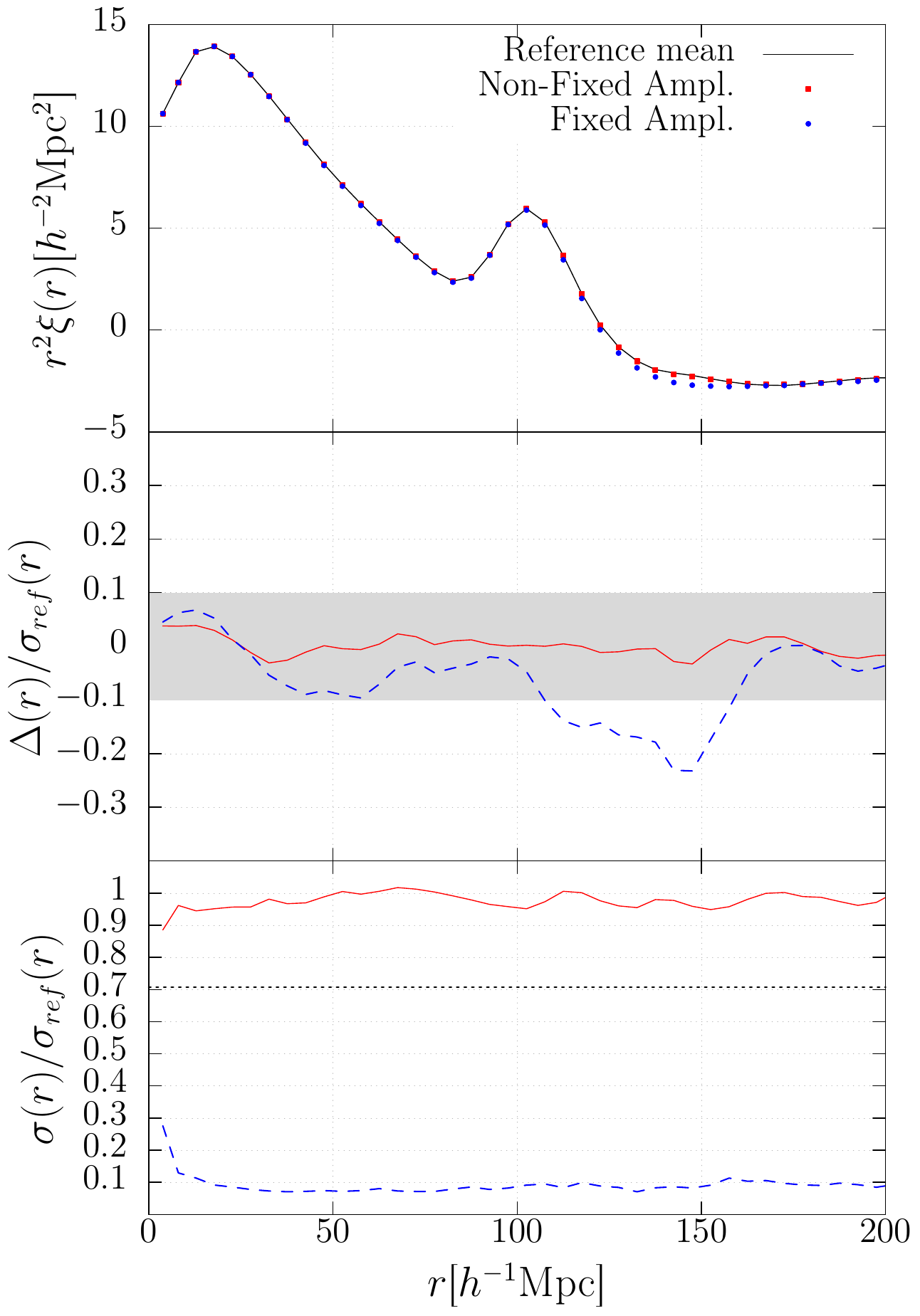}}     \subfloat[BK-particles-FastPM\label{fig:1G_DM_BK_real_z1}]{\includegraphics[width=0.33\textwidth]{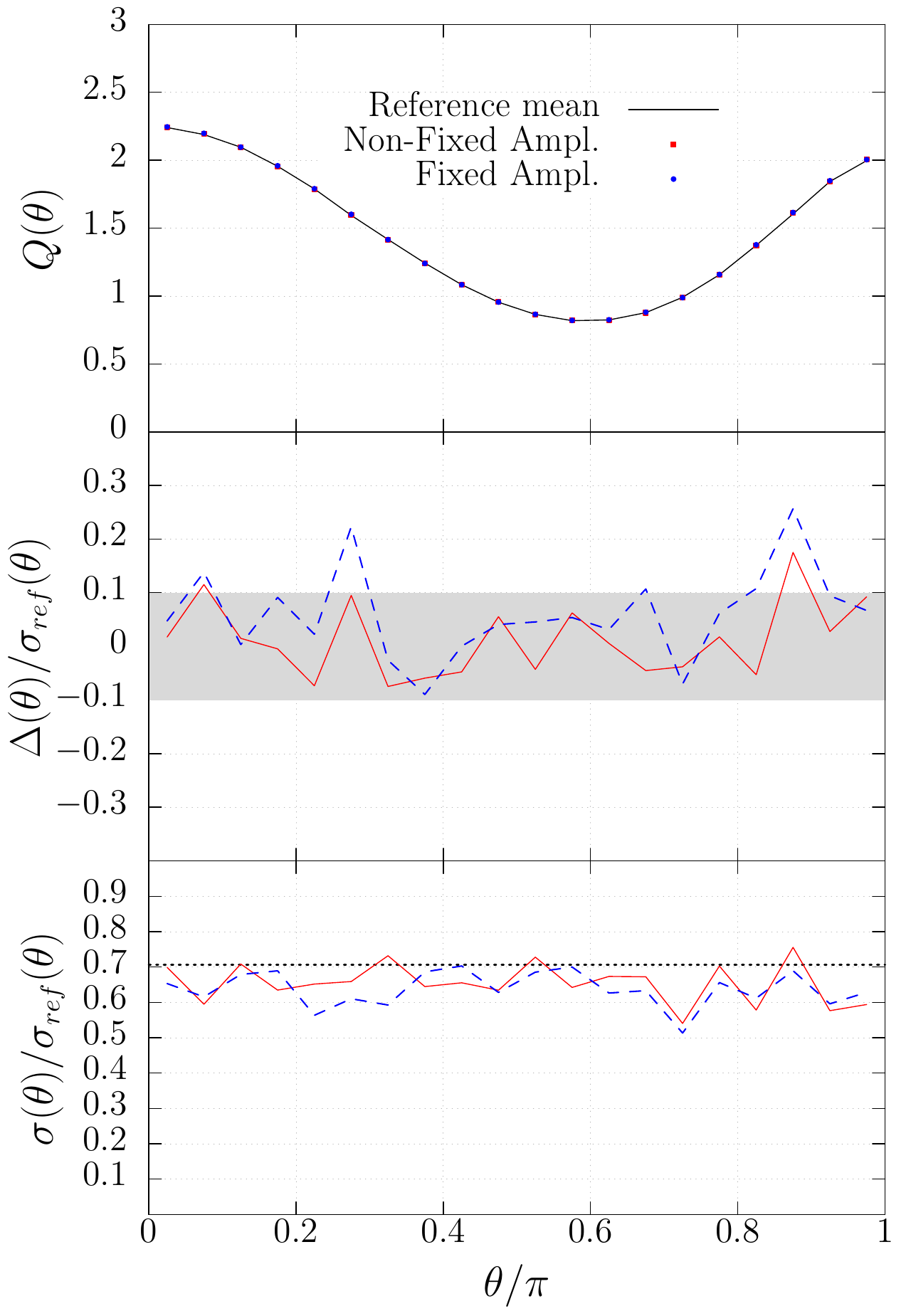}}    \caption{Performance of the Suppressed Variance Method for dark matter particles in real space in the two- (Fourier and configuration space) and three-point statistics. We show the clustering statistics for dark matter particles from FastPM runs with box size = 1 $h^{-1}$Gpc and $1024^3$ particles   at $z = 1$. 
Left, center, and right panels present the power spectra, correlation functions, and bispectra, respectively. The regular simulation set up is shown in black; the set of paired non-fixed-amplitude simulations in red, and  the set of paired-fixed-amplitude simulations in blue. 
The middle row shows the ratio between  $\Delta(k)$ and the standard deviation from the reference LR set.  
Since the uncertainty on the mean should be inverse proportional to $\sqrt{100}$, deviations between the means should be considered as unbiased if they agree within $0.1\sigma_{\rm ref}$ (gray region).
We confirm that the suppressed variance method does not introduce any statistic significant bias at any scale in the considered range.
The correlations among the data points of a correlation function are large at larger scales, so that the deviations shown in the center plot are not statistically significant either.
The bottom row shows the ratio between  the standard deviations from each paired set, $\sigma(k)$, and the reference  LR set, $\sigma_{\rm ref}(k)$. 
Since we compare the sets of paired simulation with the reference simulations, if the uncertainty is reduced by only $1/\sqrt{2}\sim0.7$, it indicates no improvement. We find significant improvement of the uncertainty from the set of paired-fixed-amplitude simulations.
However, power spectrum shows that the improvement significantly depends on the scale. 
}    \label{fig:1G_DM_real_z1} \end{figure*}

\begin{figure*}
    \subfloat[PK-halos-FastPM\label{fig:1G_HA_PK_real_z1}]{\includegraphics[width=0.33\textwidth]{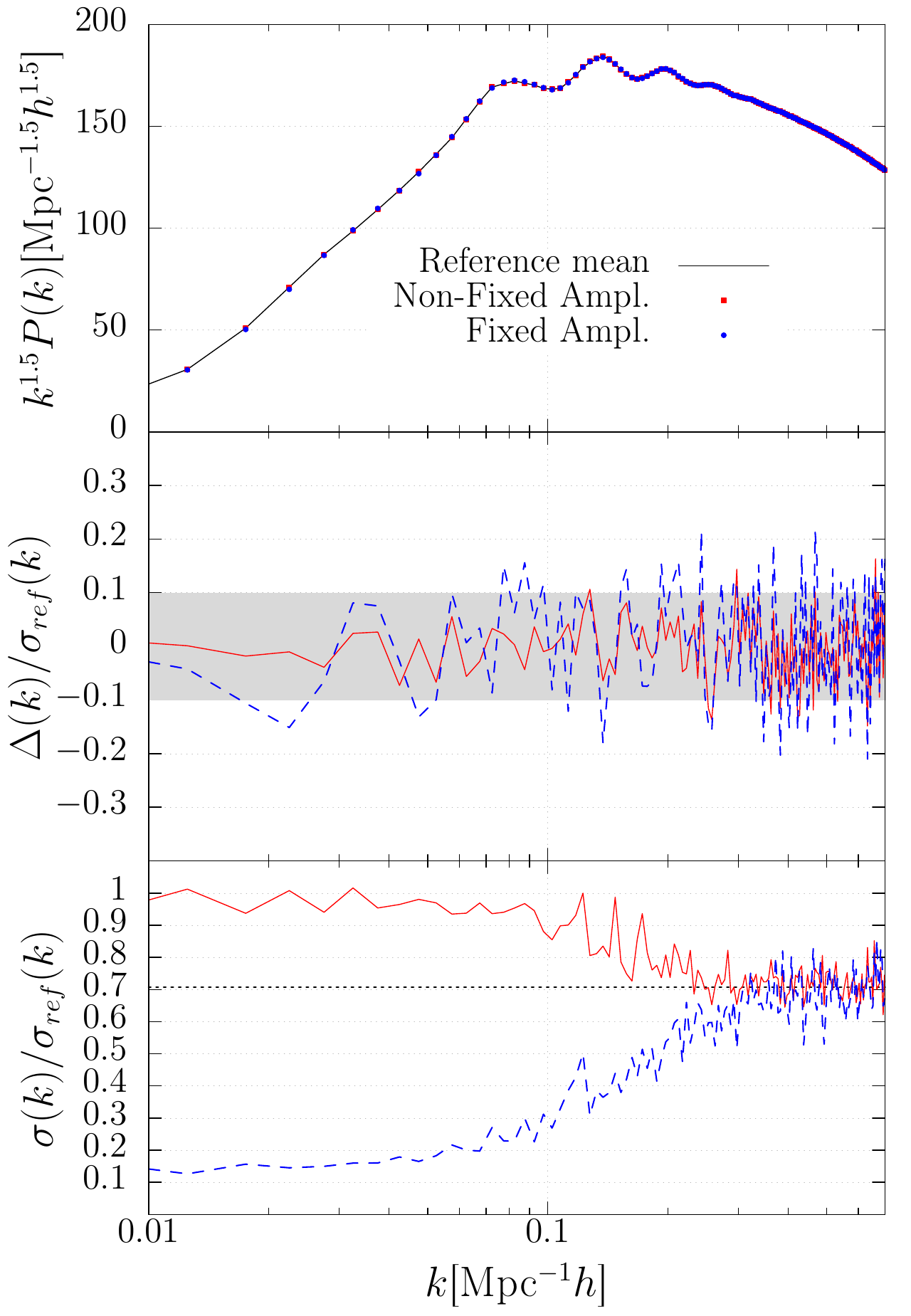}}
    \subfloat[CF-halos-FastPM\label{fig:1G_HA_CF_real_z1}]{\includegraphics[width=0.33\textwidth]{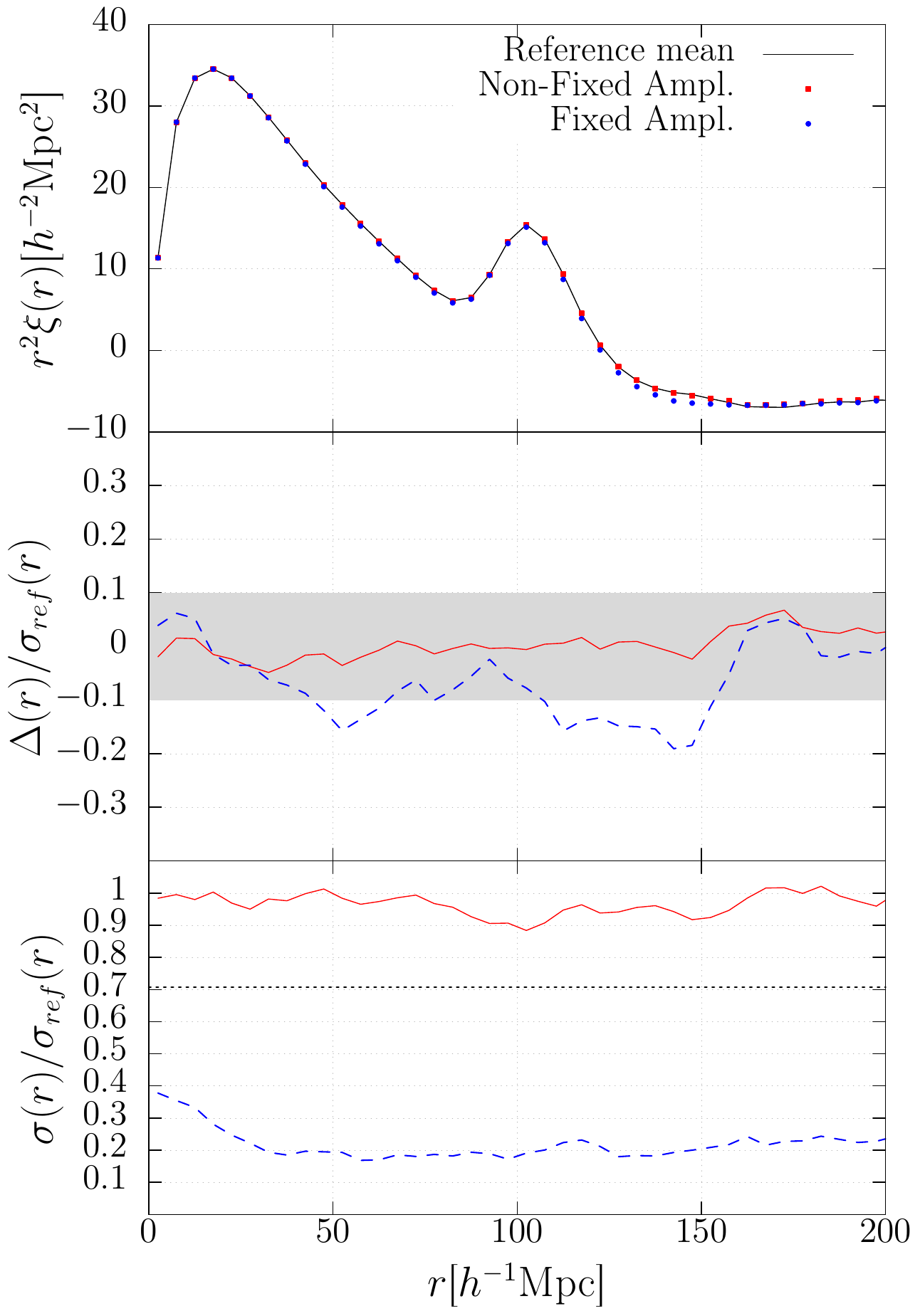}}
    \subfloat[BK-halos-FastPM\label{fig:1G_HA_BK_real_z1}]{\includegraphics[width=0.33\textwidth]{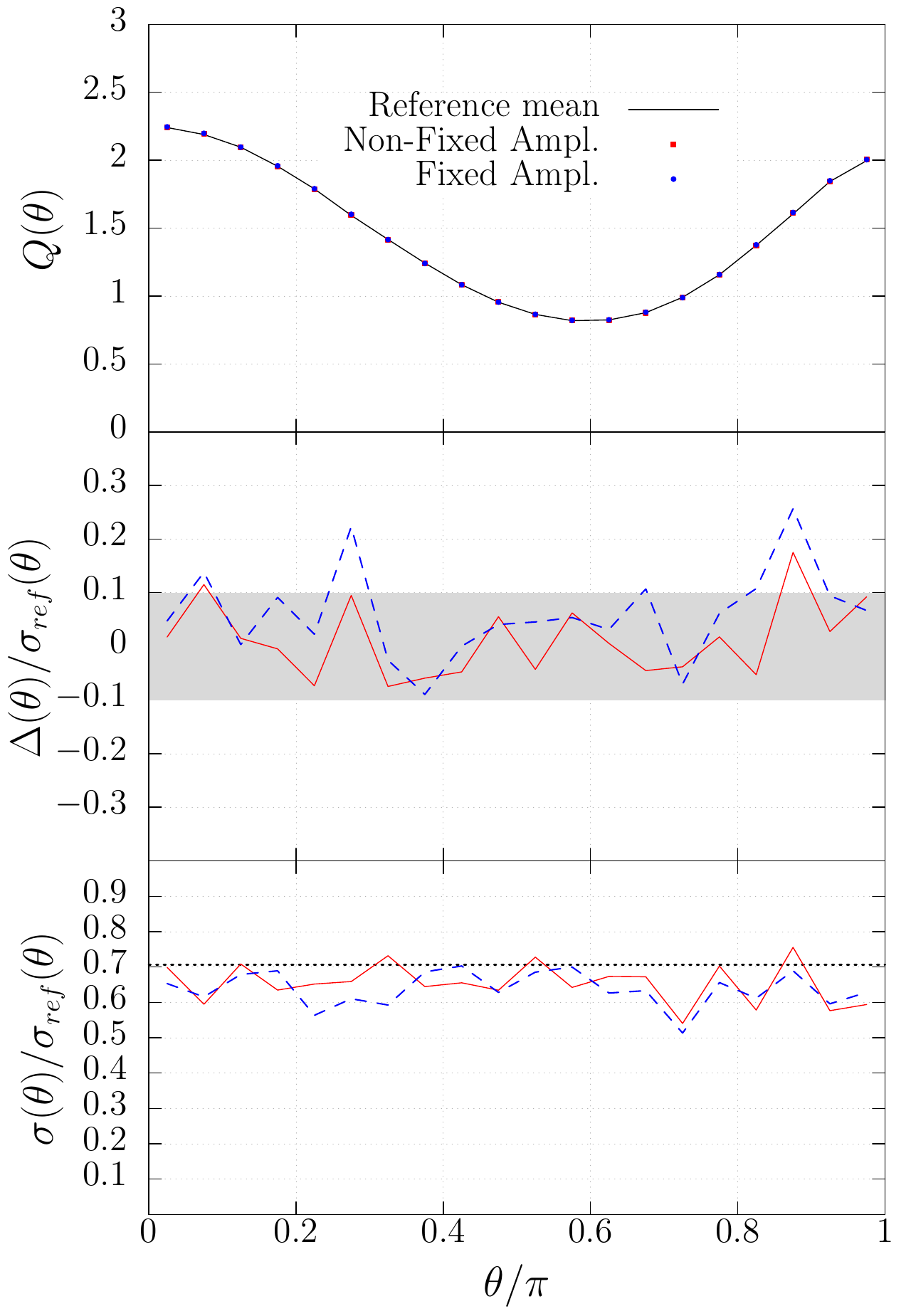}}
    \caption{Performance of the Suppressed Variance Method for dark matter haloes in real space in the two- (Fourier and configuration space) and three-point statistics.  The figure replicates Fig.~\ref{fig:1G_DM_real_z1}, but for haloes.  We do not find any bias in these measurements (see the middle panels). The improvements in the uncertainties are weaker than the ones from dark matter clustering measurements but are still significant. There is no improvement in the power spectrum for $k>0.3$Mpc$^{-1}h$ at $z=1$.}
    \label{fig:1G_HA_real_z1}
\end{figure*}

\begin{figure*}
    \subfloat[PK-mono-halos-FastPM\label{fig:1G_HA_PK_mono_z1}]{\includegraphics[width=0.4\textwidth]{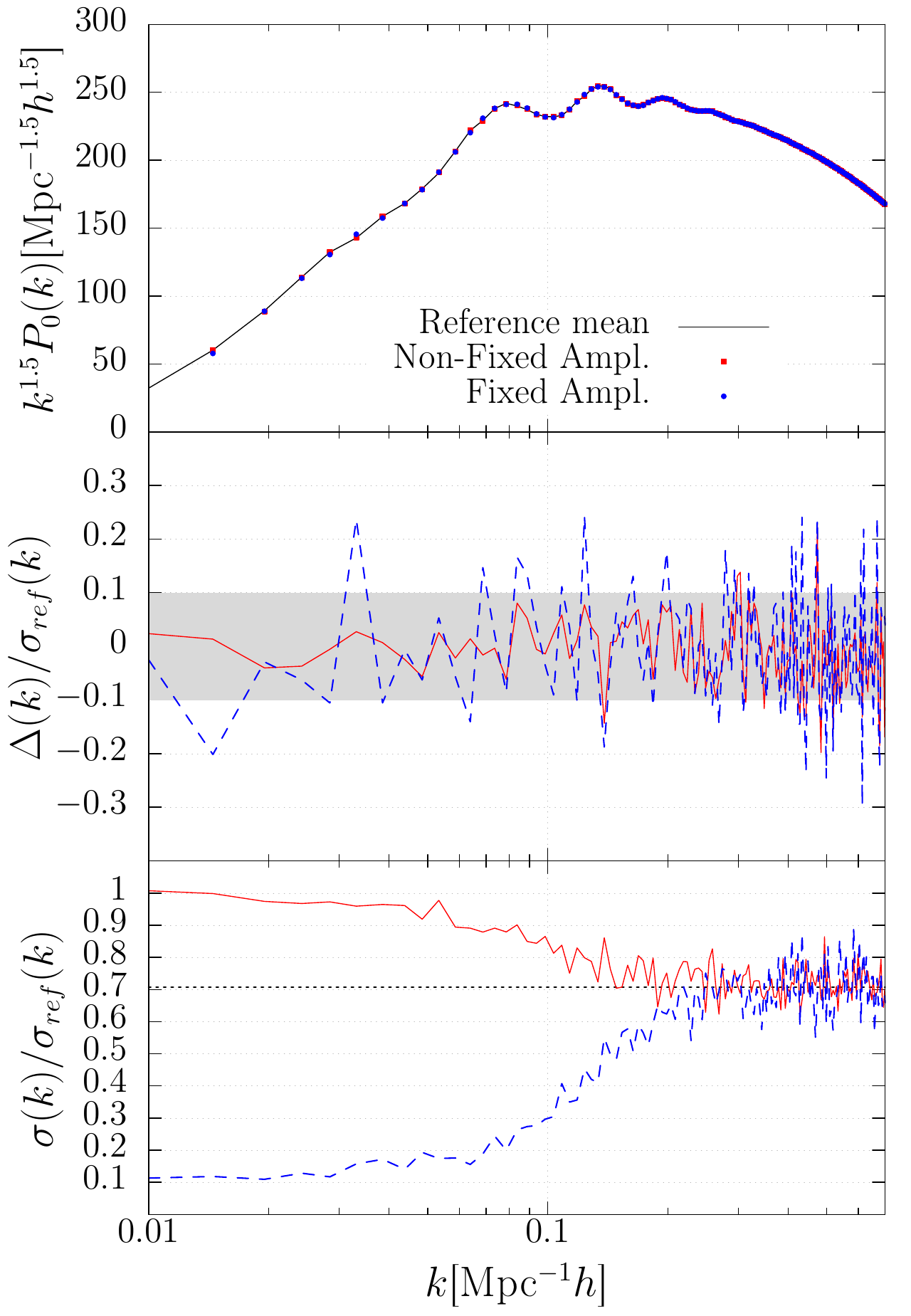}}
    \subfloat[PK-quad-halos-FastPM\label{fig:1G_HA_PK_quad_z1}]{\includegraphics[width=0.4\textwidth]{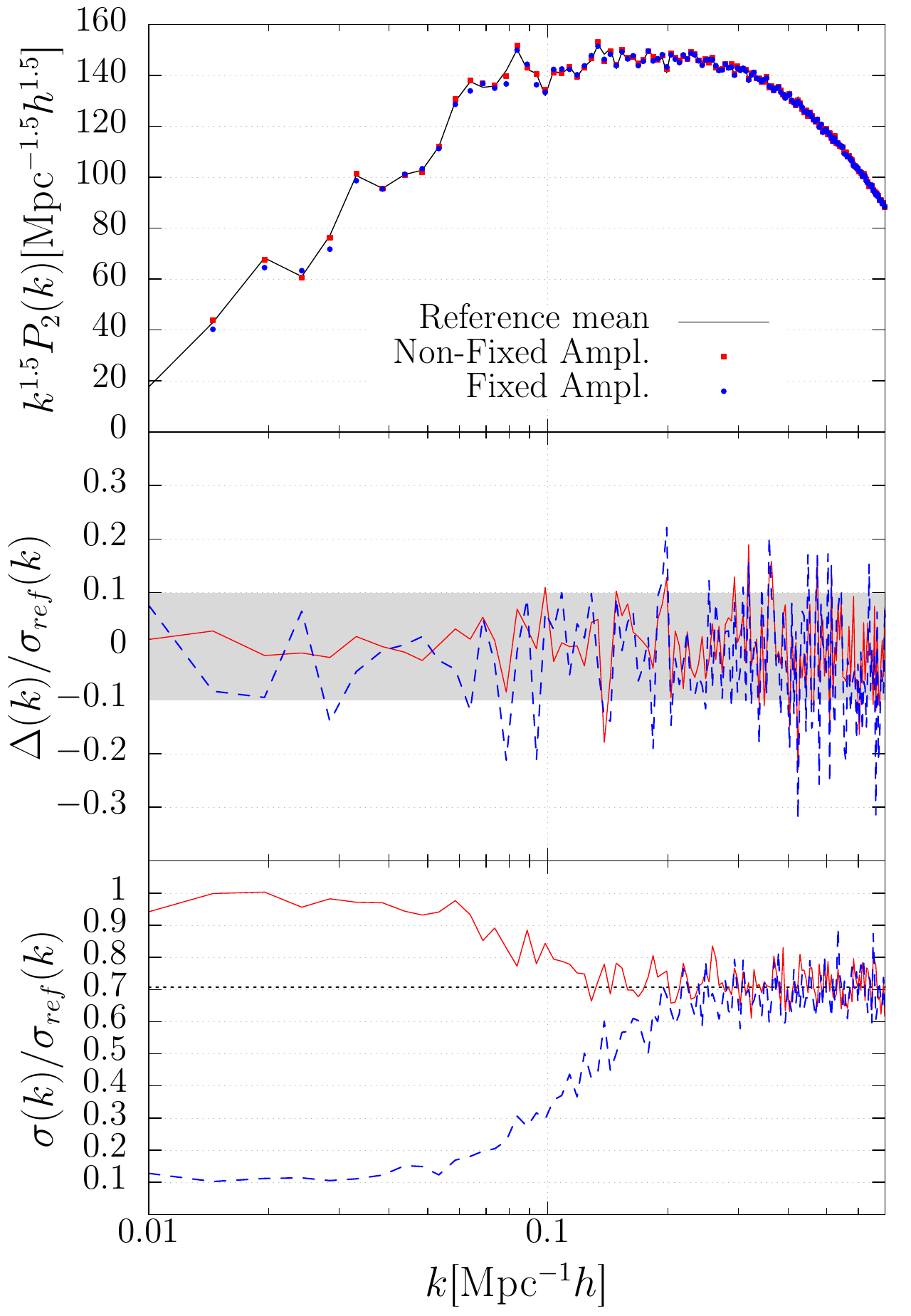}}\\
    \subfloat[CF-mono-halos-FastPM\label{fig:1G_HA_CF_mono_z1}]{\includegraphics[width=0.4\textwidth]{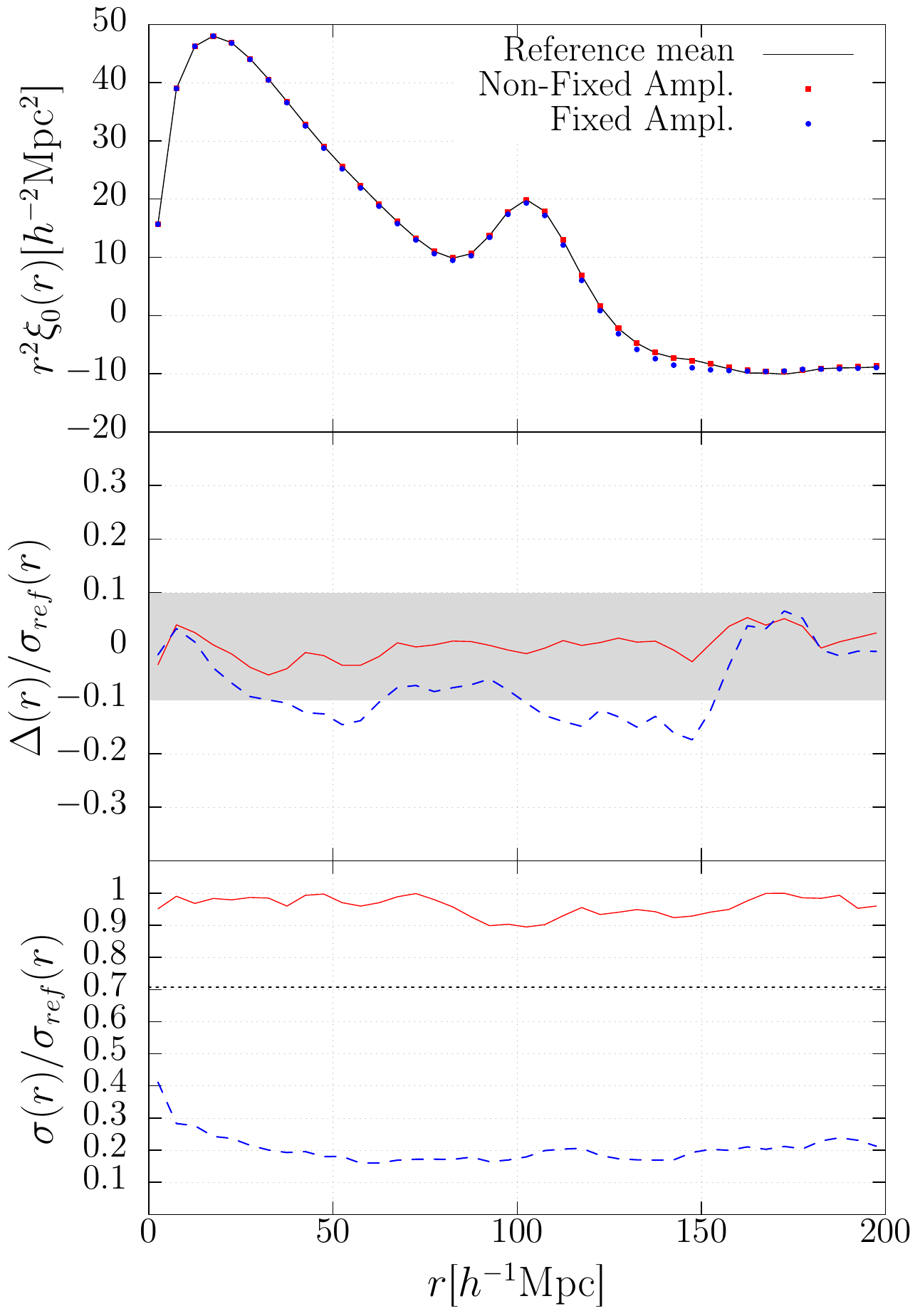}}
    \subfloat[CF-quad-halos-FastPM\label{fig:1G_HA_CF_quad_z1}]{\includegraphics[width=0.4\textwidth]
{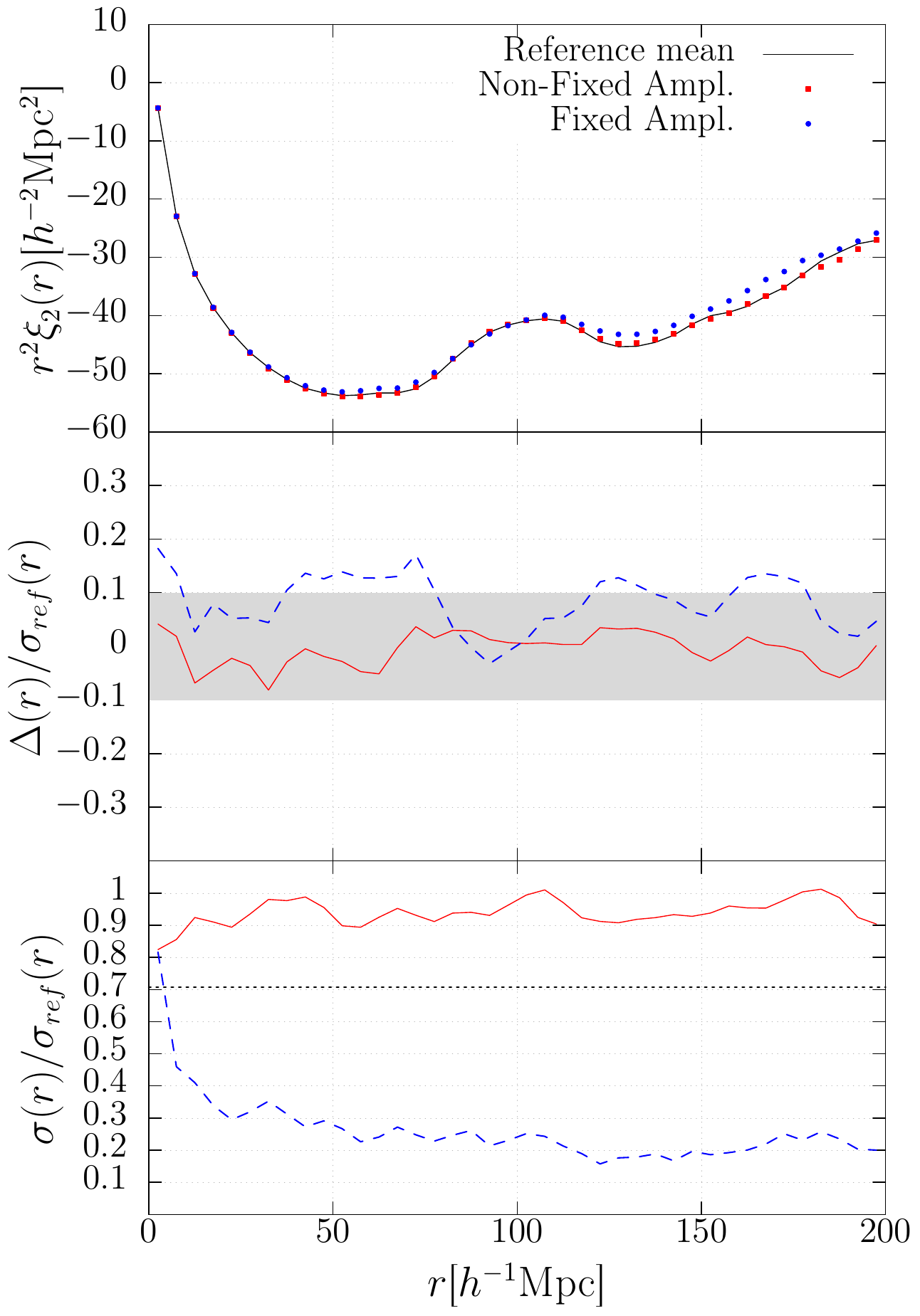}}
\caption{Performance of the Suppressed Variance Method for dark matter haloes in redshift space in the monopole and quadrupole, for both Fourier and configuration space. Panels show the monopole (left) and the quadrupole (right) in Fourier space (top) and configuration space (bottom).
The same conventions as in Fig.~\ref{fig:1G_HA_real_z1} are used. 
We do not find any bias in these measurements. The improvements are similar to those found in real space.
}
    \label{fig:1G_HA_red_z1}
\end{figure*}

\begin{figure}
\centering
\includegraphics[width=0.9\columnwidth]{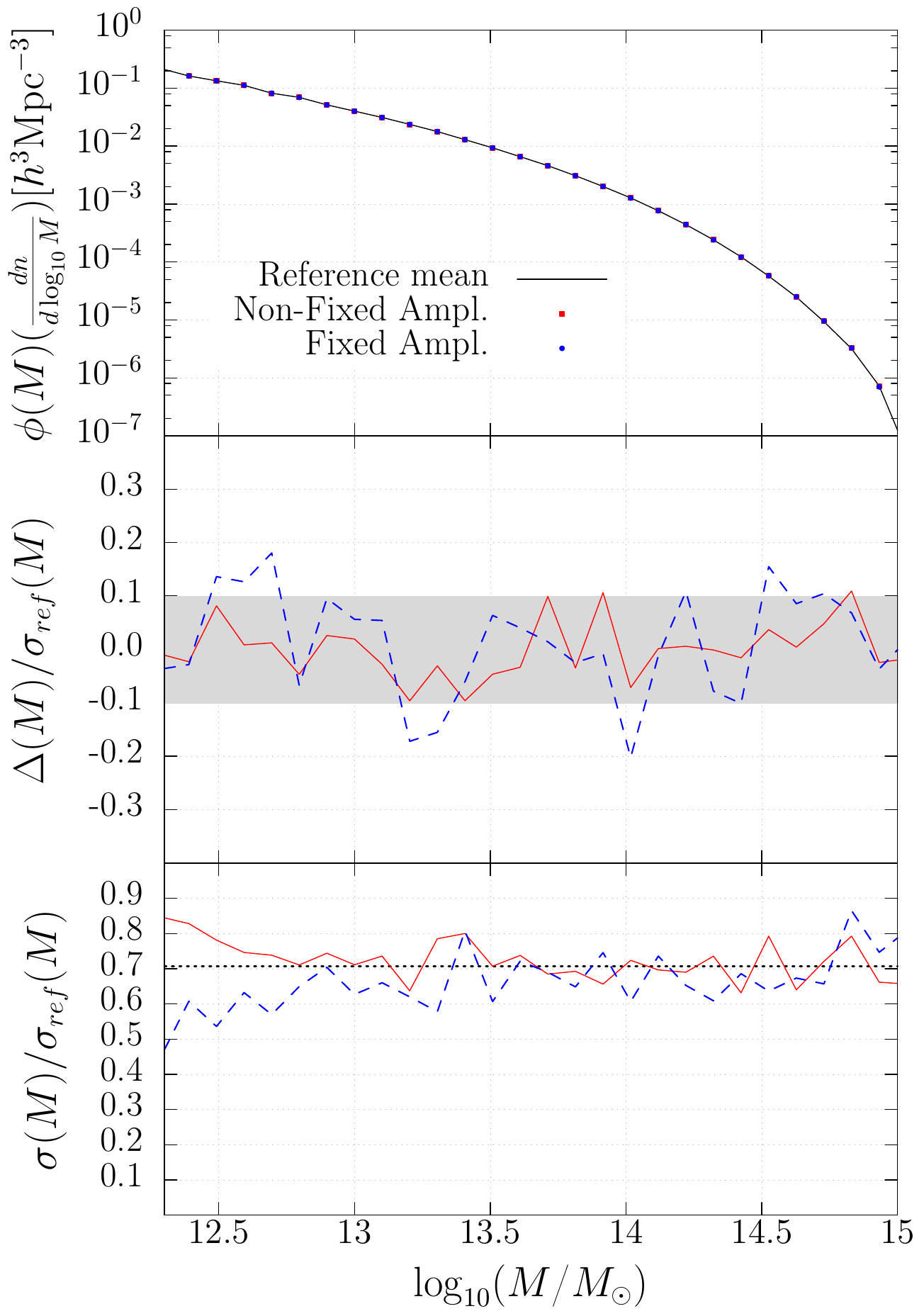}
\caption{
Halo mass functions with same conventions as in Fig.~\ref{fig:1G_DM_real_z1}.
We find no bias in the mean. We find a slight improvement in the variance below a mass of approximately $M<10^{13}\,h^{-1}\,M_\odot$.
}
    \label{fig:1G_HMF_z1}
\end{figure}

\begin{figure}
\centering
  \includegraphics[width=0.9\columnwidth]{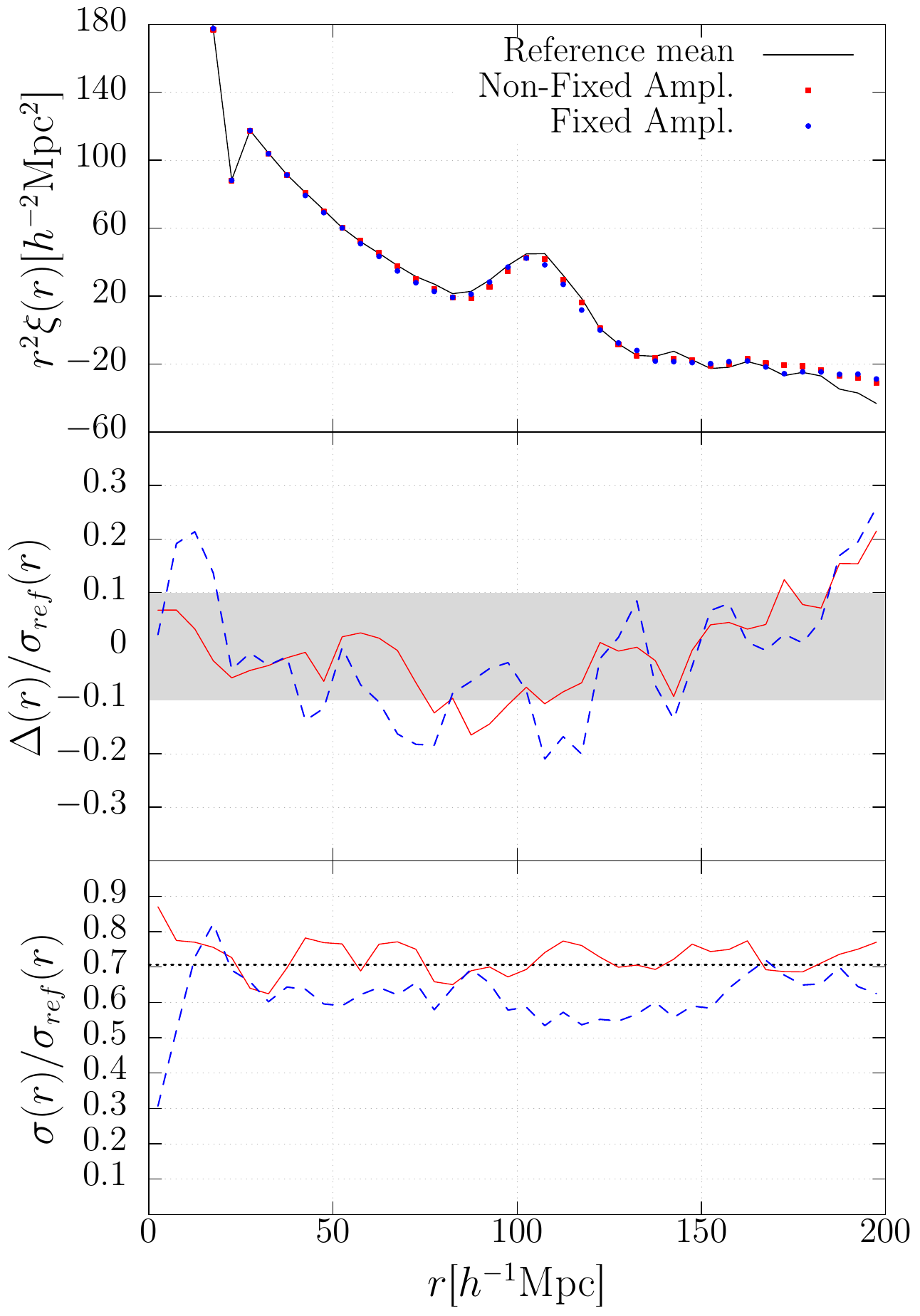}
    \caption{Void auto-correlation function selected with the radius cut 12 $h^{-1}$Mpc using the DIVE code, using the same conventions as in Fig.~\ref{fig:1G_DM_real_z1}. 
The results show no bias and a very moderate improvement in the uncertainty.}
    \label{fig:void_cf}
\end{figure}

\begin{figure}
\centering
  \includegraphics[width=0.9\columnwidth]{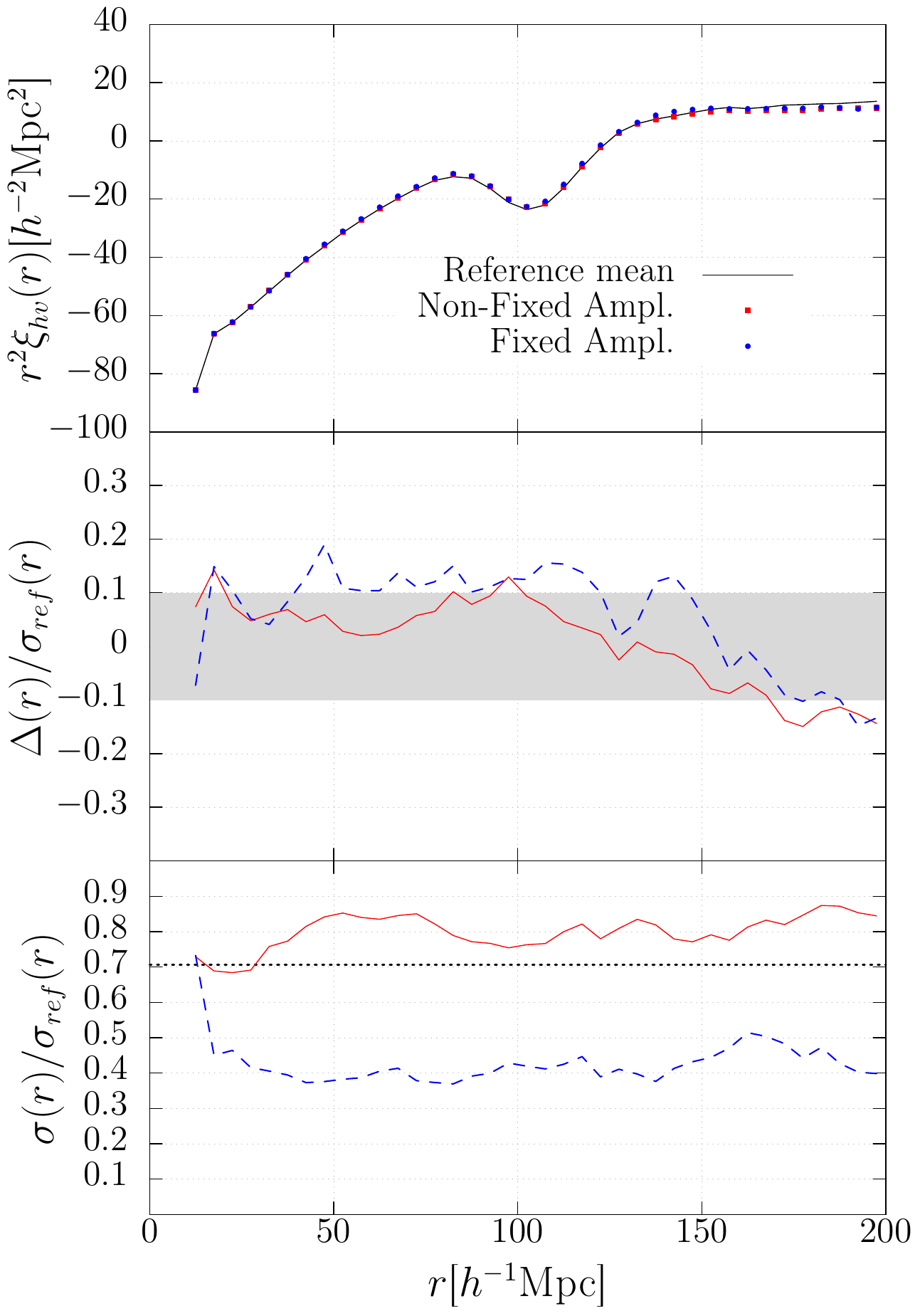}
    \caption{Same as Fig. \ref{fig:void_cf}, but for void-galaxy cross-correlation function.
This shows significant improvement in the prediction, compared to the void auto-correlation function.
}
    \label{fig:void_xcf}
\end{figure}

\subsection{Results from particle mesh simulations}

We perform an analysis of several different clustering statistics, including the power spectrum (PK), correlation function (CF), and bispectrum (BK) (as defined for instance in \citealt{Chuang:2016uuz}) using dark matter particles and halos based on the set of fast particle mesh simulations.  We demonstrate below that there are no systematic biases using the SVM, and that the variance is indeed greatly reduced in two-point statistics over the scales of interest to BAO and RSD analyses.

We quantify the uncertainties and biases in these measurements through the standard deviation of the reference simulations: $\sigma_{\rm ref}(k)$, and the  deviation of the mean with respect to the reference mean: $\Delta(k)$.
The top panels in Figs. \ref{fig:1G_DM_real_z1} and \ref{fig:1G_HA_real_z1}  show the original comparison of the clustering statistics; the middle panels show the comparison of the mean normalized by the uncertainty of the reference low resolution (LR) 100 boxes. Since the uncertainty on the mean should be inversely proportional to $\sqrt{100}$, deviations between the means should be considered as unbiased if they agree within $0.1\sigma_{\rm ref}$.
This study is performed for dark matter particles, halos, and cosmic voids, as we discuss below.

\subsubsection{dark matter particles}

The largest suppression of variance is obtained for the dark matter distribution. Fig. \ref{fig:1G_DM_real_z1} shows the comparison of the dark matter particle clustering measurements from FastPM runs   with 1 $h^{-1}$Gpc side boxes and $1024^3$ particles at $z = 1$, including the power spectrum (PK), correlation function (CF), and bispectrum (BK). In each plot, the top panel shows the clustering measurements of the reference set, the set of paired simulations (non-fixed-amplitude), and the set of paired-fixed-amplitude simulations; the middle panel shows the difference of the mean from each paired set and the reference one divided by the standard deviation from the reference set; the bottom panel shows the ratios of the standard deviations from each paired set and the reference one. From these calculations we confirm that the suppressed variance method does not introduce significant bias at any scale in the considered range.  Since the paired simulations have twice volume of the reference simulation, there is no improvement if the uncertainty of the paired simulations is larger than or equal to $1/\sqrt{2}\sim 0.7$ of that measured from the reference simulations.   We find that the improvement depends on scale. In the case of the power spectrum, we find that the improvement is significant at small $k$ (large scales) but small at large $k$. If one considers for instance $k>0.3\,h$ Mpc$^{-1}$, our results indicate that the variance at large $k$ is dominated by higher-order mode coupling terms. Interestingly, the improvement in the correlation function variance is nearly constant ($0.1\sigma_{\rm ref}$) over the range for $r>10\,h^{-1}$Mpc. We do not find any improvement in the bispectrum with triangle configurations of $k_1=0.1$ and $k_2=0.2\,h$ Mpc$^{-1}$). Since we are interested in scales relevant to BAO and RSD analysis, this study goes further into the nonlinear regime than did the study of \cite{Angulo:2016hjd}, which found models improvement for the bispectrum variance at scales that are significantly more linear.
The data products made available at the project website will enable a deeper investigation of these effects.

\subsubsection{dark matter haloes}

Dark matter haloes show the same qualitative results as for the dark matter particles; quantitatively the suppression of variance is more modest.
This is illustrated in Fig. \ref{fig:1G_HA_real_z1}, which compares halo clustering measurements, analogous to Fig. \ref{fig:1G_DM_real_z1}.
From this we conclude that the suppression of variance method is also not biased for dark matter haloes catalogs at any scale over the range considered. As in the case of dark matter clustering measurements, the improvement in the variance depends on the scale.  
In this case for the PK and CF at large scales, the ratios of the uncertainties (the bottom panels) can be as small as 0.2 or less, corresponding to an effective simulation volume of more than 25 ($h^{-1}$Mpc)$^3$. As in the dark matter particle case, we do not find improvement for the BK.

The halo population is a biased subset of all matter. Haloes' masses and positions are sensitive to small-scale fluctuations in the initial density field, an effect often referred to as stochasticity. This stochasticity explains the difference in the results with respect to the performance of the dark matter distribution traced by particles.


The results for redshift-space halo clustering including monopoles and quadrupoles in configuration and Fourier space show a similar performance to the real-space measurements, as shown in Fig. \ref{fig:1G_HA_red_z1}. We use the same definition of the multipole expansion as in \cite{Chuang:2016uuz}.

\subsubsection{halo mass function}

As another relevant statistic, we investigate the halo mass function, shown for $z=1$ in Fig. \ref{fig:1G_HMF_z1}. We find no bias in the mean, and only a slight improvement in the variance below a mass of approximately $M<10^{13}\,h^{-1}\,M_\odot$. 
We further check this in the additional set of FastPM boxes with smaller box size but higher resolution and confirm the improvement of the mass function in the lower mass bins.  Further tests are shown at \url{http://www.unitsims.org}. In that supplementary material we also show that the suppression of variance is more effective a) towards increasing redshifts, as structure formation becomes more linear; and b) for lower mass cuts, as the higher mass populations suffer more from stochasticity (see also Section \ref{sec:gadget} for another representation of these trends).

\subsubsection{void clustering}

We now consider cosmic voids,  focusing on the well-defined convention used in the void finder code DIVE \citep{Zhao:2015ecx} that considers voids as empty spheres constrained by quartets of galaxies. This definition has proved useful to study the troughs of the density field, i.e. the clustering within cosmic voids, and to obtain improved measurements of the BAO signature (see \citealt{Kitaura:2015ubm,Liang:2015oqc,Zhao:2018hoo}).
Cosmic voids in fact are measures of the higher-order statistics of the galaxy distribution (see above mentioned papers and references therein), and are therefore interesting to study the performance of suppressed variance methods.
As expected from the BK result, the auto-correlation function of the voids, shown in Fig. \ref{fig:void_cf}, shows a very moderate improvement in the uncertainty and no bias. 
However, the cross-correlation functions between halos and voids present significant improvements, shown in Fig. \ref{fig:void_xcf}.

\section{Application of SVM for clustering analysis from galaxy surveys}
\label{sec:gadget}
We have demonstrated in the previous section that the suppressed variance method does not introduce any bias, and significantly reduces uncertainty in the two-point statistics. We now describe our first two pairs of high-resolution full $N$-body simulations aimed at the analysis of ELG and LRG data from DESI- and Euclid-like surveys.

\begin{figure*}
\centering
\includegraphics[width=1.9\columnwidth]{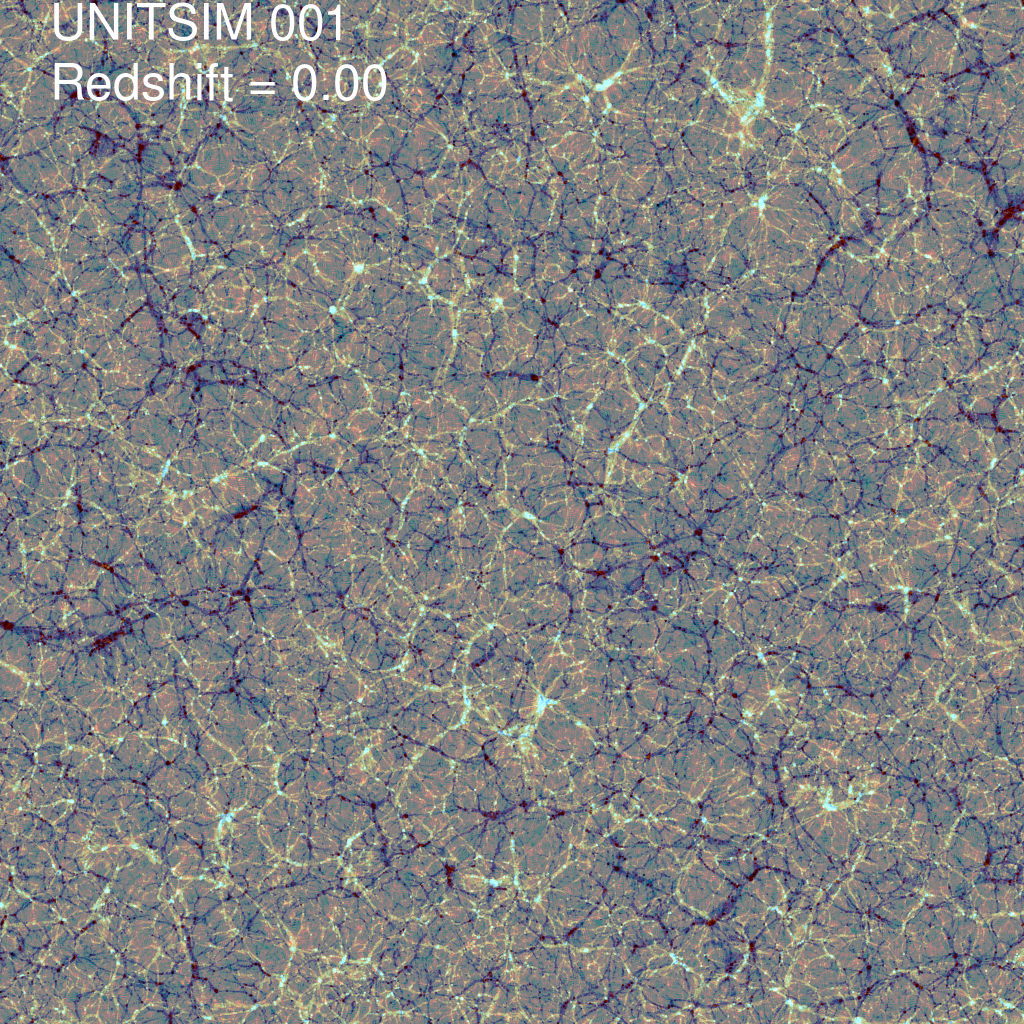}
\caption{
A slice of 500 $\times$ 500 $h^{-1}$Mpc and 0.5 $h^{-1}$Mpc thickness from the density fields of a pair of GADGET simulations. We show one simulation in bright color and one in dark color. One can see that the overdensity regions (e.g. knots) in one simulation are the underdensity regions (e.g. voids) in the other one.
}
    \label{fig:gadget_density}
\end{figure*}

\begin{figure*}
\subfloat[PK-halo-Gadget\label{fig:Gad_HA_PK_z1}]{\includegraphics[width=0.33\textwidth]{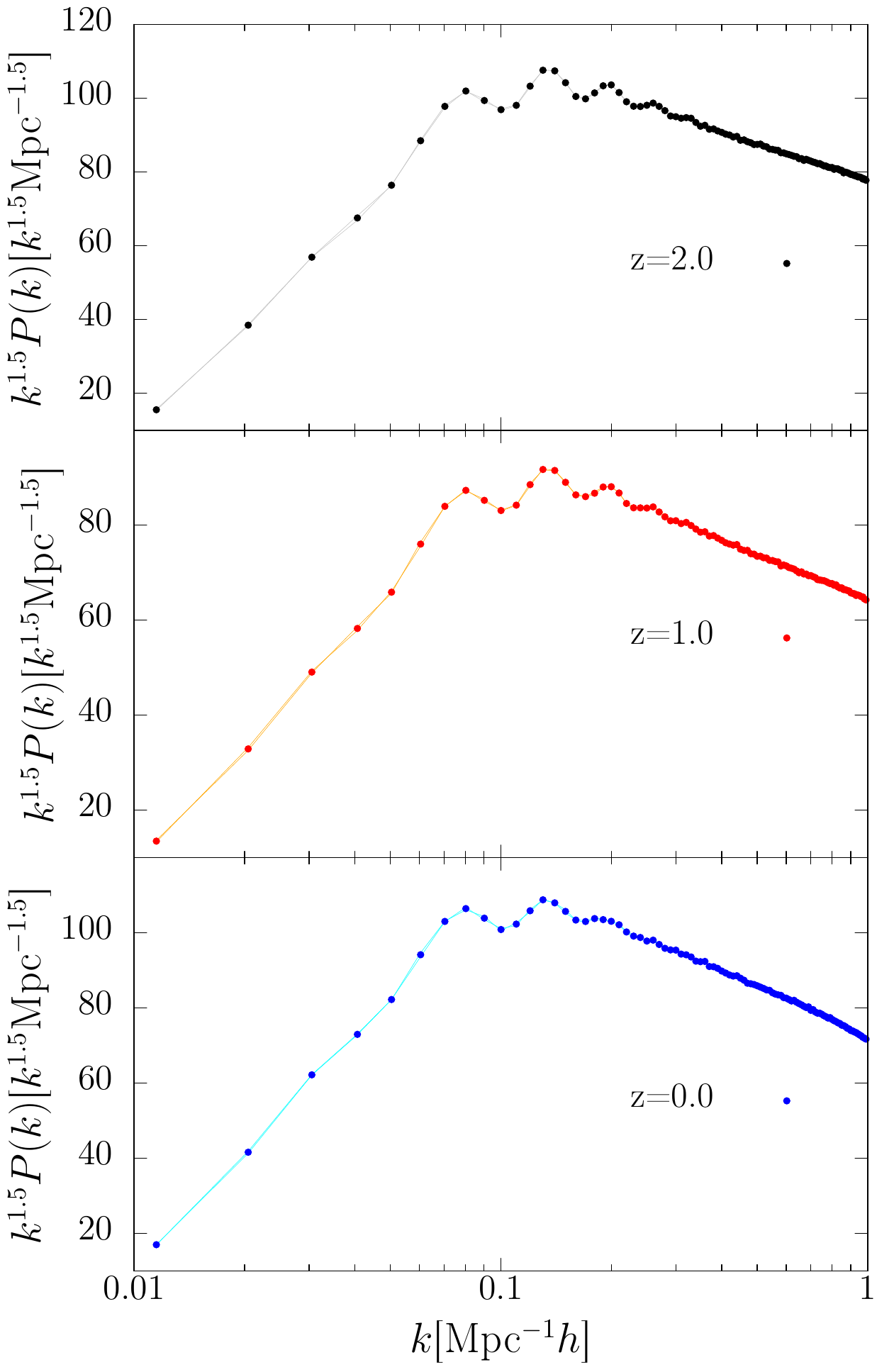}}
\subfloat[CF-halo-Gadget\label{fig:Gad_HA_CF_z1}] {\includegraphics[width=0.33\textwidth]{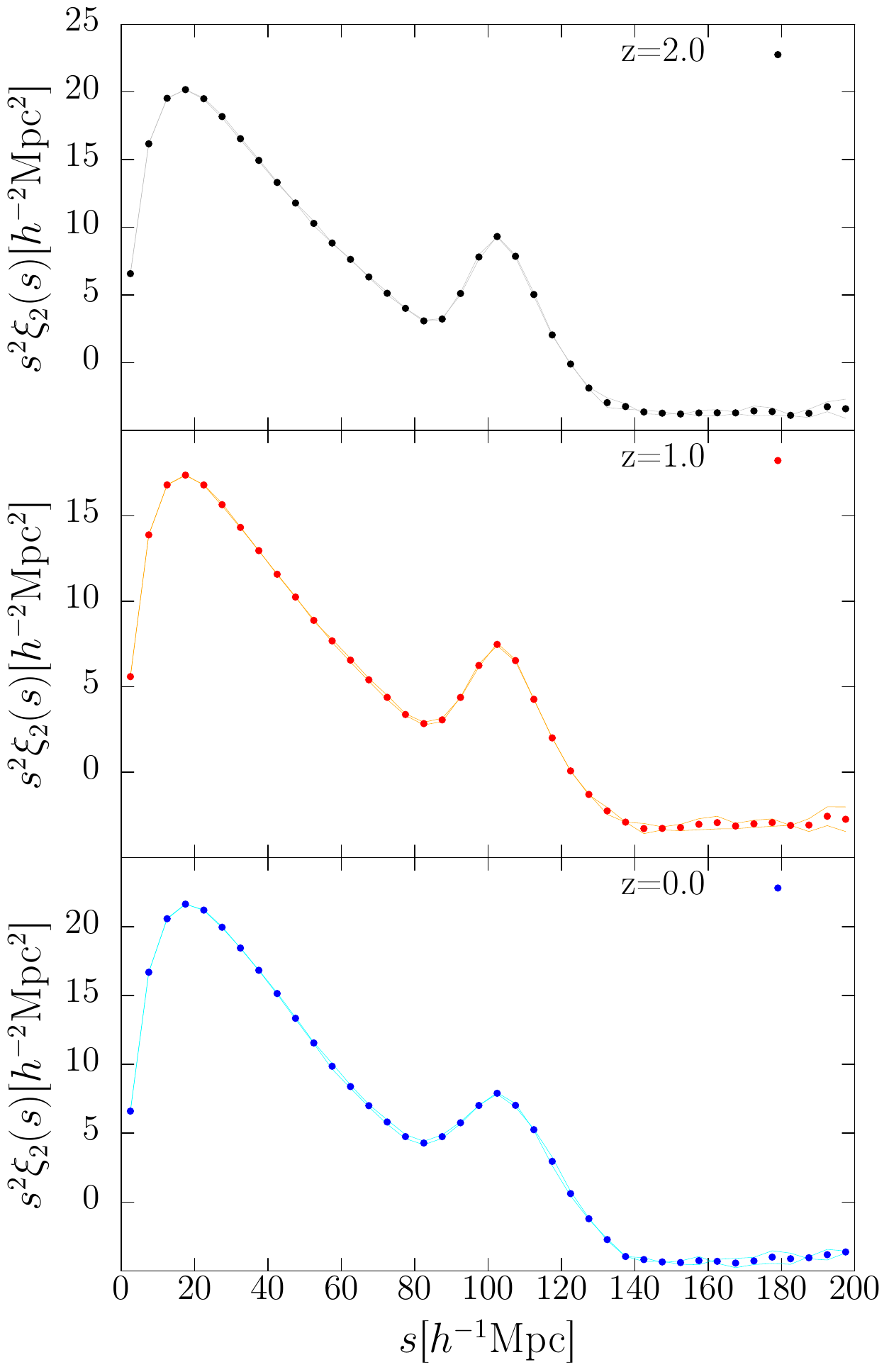}}
\subfloat[HMF-halo-Gadget\label{fig:Gad_HA_HMF_z1}]{\includegraphics[width=0.33\textwidth]{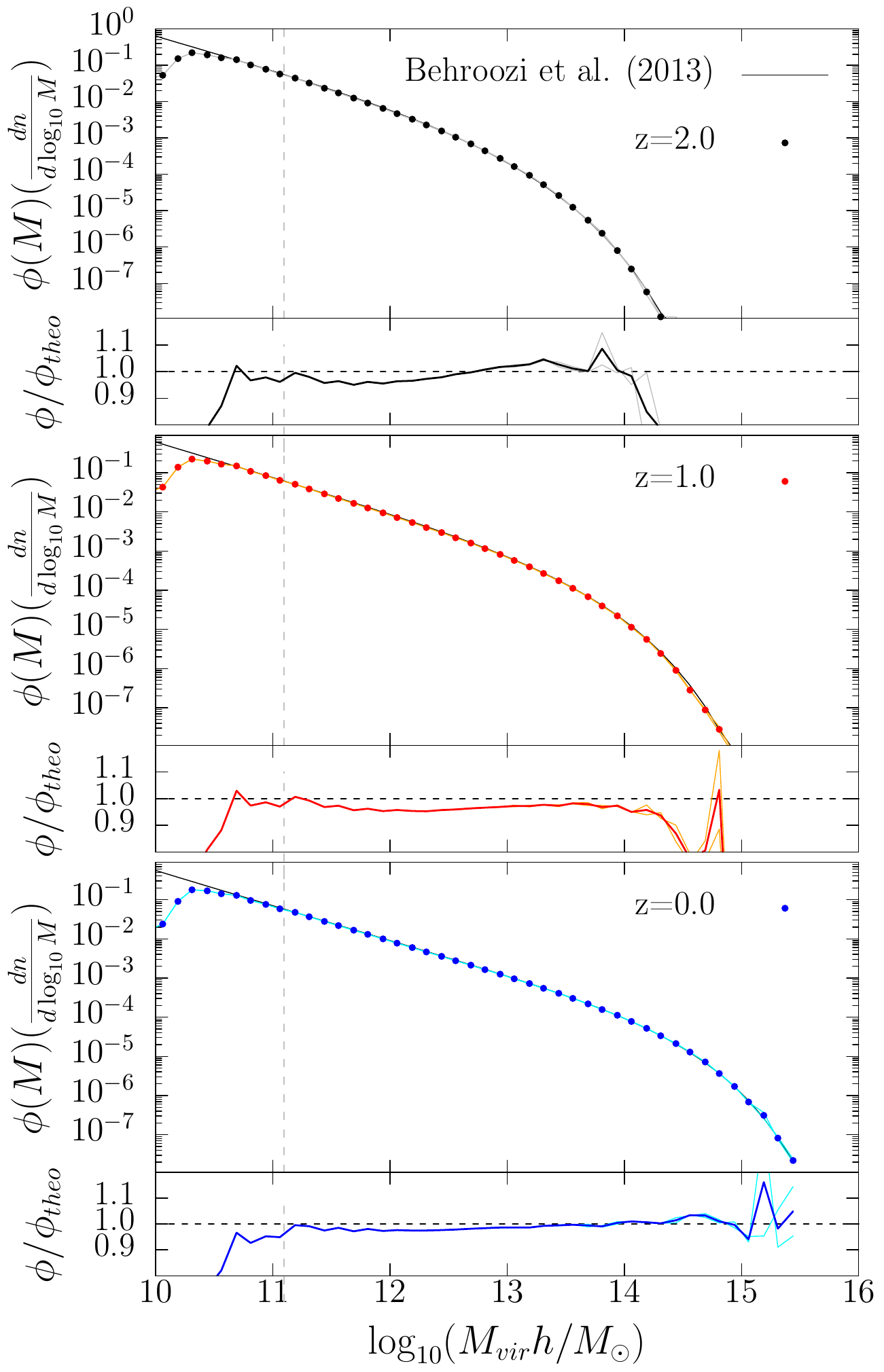}}
    \caption{ 
Power spectrum, correlation function, and halo mass function measurements from full $N$-body simulations with the SVM.
SVM reduces the variances significantly so that the measurements are very smooth.}
    \label{fig:gadget-clustering}
\end{figure*}

\subsection{Setup}

We use the $N$-body code {\sc Gadget} \citep{2005MNRAS.364.1105S}, a full MPI parallel code that uses Particle-Mesh (PM) + Tree algorithms to  compute the Newtonian forces between the dark matter particles by splitting the gravitational force into a long-range term (computed through the PM method) and a short-range term taken from the nearest neighbors, using a Tree method to categorize the particles according to their relative distances.  This code makes use of the public software library FFTW for parallel Fast-Fourier transforms and the GNU Scientific Library (GSL). We are using a non-public version of the {\sc Gadget} code, {\sc L-Gadget},  that is highly optimized for large-volume simulations with a cubic domain decomposition and an efficient use of internal memory. This code has been extensively used to produce large-volume simulations with billions to hundreds of billions of particles, including the Multidark simulation suite (see \url{http://www.multidark.org}) and the Millennium series of simulations (including the Millennium XXL with more than 300 billion particles).

The paired initial conditions with fixed amplitude are generated using second order Lagrangian perturbation theory with FastPM \citep{Feng:2016yqz}. We use the same cosmology as the FastPM simulations generated for this study (see Section \ref{sec:fastpm}). The box size is $1 h^{-1} \rm{Gpc}$ and the simulation is started at $a\equiv 1/(1+z)=0.01$ ($z=99$). The number of particles is $4096^3$, giving a particle mass is $\sim 1.2\times10^9$ $h^{-1}$M$_{\odot}$. 
A slice of 500 $\times$ 500 $h^{-1}$Mpc and 0.5 $h^{-1}$Mpc thickness from the density fields of a pair of {\sc Gadget} simulations. We show one simulation in bright color and one in dark color. One can see that the overdensity regions (e.g. knots) in one simulation coorespond to the underdensity regions (e.g. voids) in the other one.
We use the halo finder code {\sc Rockstar} \citep{Behroozi13} to identify haloes and compute their merging histories using the {\sc Consistent Trees} software \citep{consistent_trees}.

\subsection{Results from full $N$-body simulations}
Below we present tests based on two sets of pairs of full $N$-body simulations. The resolution of these Gpc scale simulations has been chosen to match the resolution of our small-volume particle mesh simulations using the FastPM code. The halo catalogs were generated using a minimum halo mass of $1.2\times 10^{11}$ $h^{-1}$M$_{\odot}$ (at this limit the mass function is quite complete, as shown in the right-most panel of Fig. \ref{fig:gadget-clustering}). This permits us to assess the improvement in the statistics from the FastPM simulations with the 250 $h^{-1}$Mpc box size and a  mesh of $1024^3$. We show the power spectrum, correlation function, and halo mass function measurements from our fixed-amplitude-paired Gadget $N$-body simulations in Fig. \ref{fig:gadget-clustering}, which turn out to be remarkably smooth for the different redshift snapshots.  We explore robust statistical measures in the next section to further assess the quality of the simulations.

\begin{figure*}
\centering
\includegraphics[width=1\textwidth]{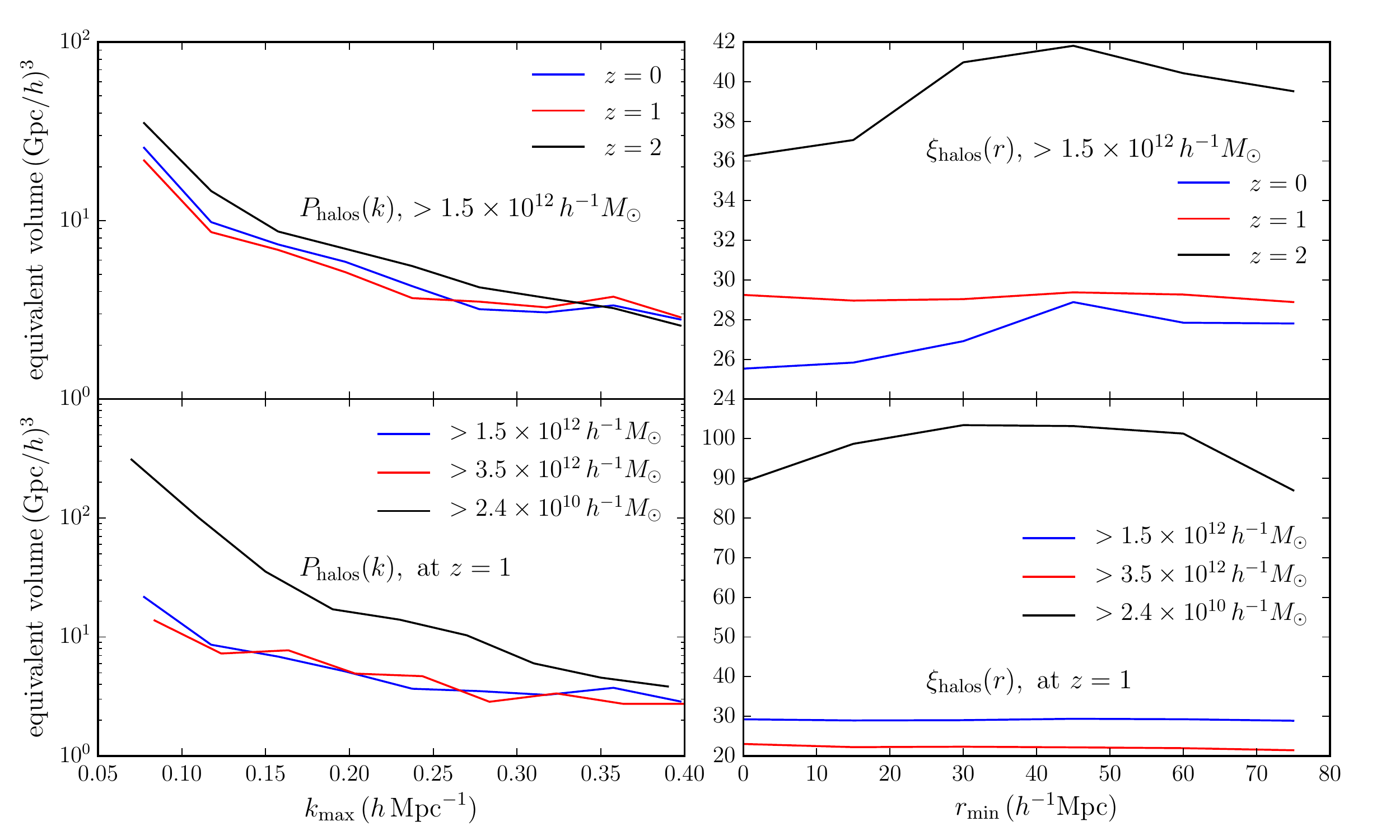}\\
\includegraphics[width=1\textwidth]{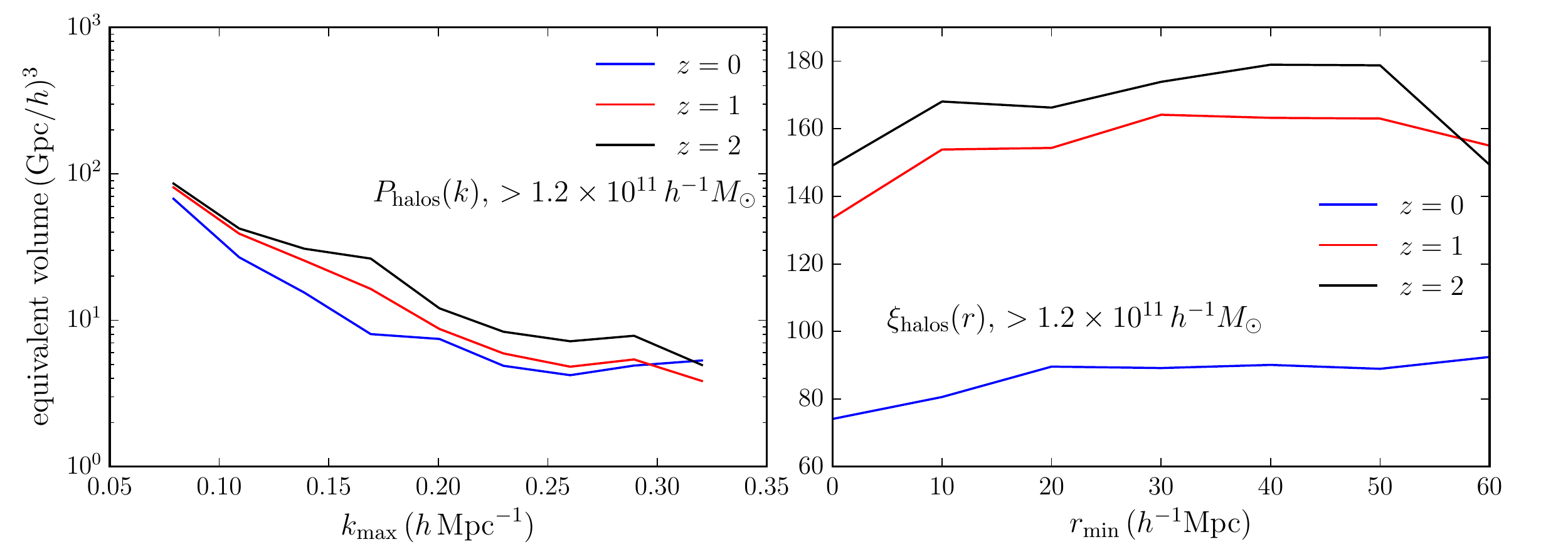}\caption{Equivalent volume study of the SVM for different redshifts and mass cuts.
The upper four panels show the equivalent volumes of the catalogues from one pair of 1 $h^{-1}$Gpc boxes.
The lower two panels show the equivalent volumes of the catalogues from two pairs of 1 $h^{-1}$Gpc boxes based on our high-resolution full $N$-body simulations.
The results shown in the left panels are in Fourier space (varying $k_{max}$ with fixed $k_{min}=0$) and those shown in the right panels are in configuration space (varying $r_{min}$ with fixed $r_{max}=120$).
}\label{fig:improve_100par}\end{figure*}

\subsubsection{Estimator quantifying the improvement in SVM}
\label{sec:improve}
Thus far, we have shown the improvements at different scales for different clustering statistics. However, in a practical cosmological analysis (see e.g. \citealt{Chuang:2016uuz}), we use a specific scale range (e.g. $40<r<200\,h^{-1}$ Mpc in configuration space or $0.02<k<0.2 \,h$ Mpc$^{-1}$ in Fourier space), so the improvement should be determined by the whole range. To quantitatively assess the improvement, we adopt the Fisher information matrix formalism evaluating the improvement of the constraining power on a given cosmological parameter by performing the analysis within a certain scale range. In this approach, the uncertainty of a given cosmological parameter ($\theta$) is defined as 
\begin{eqnarray}
\label{sq:salmon}
{\rm Var}(\theta) 
&=& (F^{-1})_{\theta, \theta}\nonumber \\
&=& \left \langle \frac{-\partial^2 \ln {\rm P}(\theta)}{\partial \theta^2} \right \rangle^{-1}\nonumber \\
&\propto& \Bigg \langle -\frac{\partial ^2 \left( {\rm Tr [\log C ]}+ \sum_{ij} f_i {\rm Tr C}^{-1}_{ij} f_j \right)}{\partial \theta^2}\Bigg \rangle ^{-1}
\end{eqnarray}
where P($\theta$) is the posterior, and $\langle ... \rangle$ denotes the expectation value. $C$ is the covariance matrix of some measurement $f$ (e.g. PK or CF), and $i$, $j$ are the indices of the elements, i.e. $C_{ij} = \langle f_i f_j\rangle $ \citep{Dodelson2003book}.
We have assumed a flat prior on the parameter $\theta$ and a Gaussian likelihood. For simplicity, we further assume that all the measured data points within the scale range of interest have the same sensitivity to the parameter $\theta$, i.e. $\partial {\rm f_i}/\partial \theta = \partial {\rm f_j}/\partial \theta, \, \forall i,j$ . With these  assumptions, the uncertainty of the parameter $\theta$ can be related to the covariance matrix of the data vector via the following equation, 

\begin{equation}\label{eq:Hamachi}
 {\rm Var(\theta)} = A\left(\sum_{i,j} \left(C^{-1}\right)_{ij}\right)^{-1}, 
\end{equation}
where $i,j$ go through all the data points within the scale range of interest, and $A$ is assumed to be a constant involving terms $(\partial { f}/\partial \theta)^2$.

We then quantify the covariance matrix of the data vector from the simulation. We note that in order to perform the cosmological analysis, one has to account for two types of uncertainties. The first one is the theoretical uncertainty, represented by the theoretical covariance, $C_{\rm theory}$, which is driven by the standard error on the statistic, $f$ in the simulations used to validate the models. The second one is the observational uncertainties encoded in the covariance matrix $C_{\rm obs}$. For a galaxy survey, for example, this would include sample variance on large scales, and stochasticity on small scales. The total covariance matrix is given by the sum of the individual ones, i.e.
\begin{equation}
    C = C_{\rm theo} + C_{\rm obs} \, .
\end{equation}
The reasonable assumption here is that there are no cross-covariances between the two.

The theoretical covariance matrix, $C_{\rm theo}$, can be calculated from the simulations used for validating the models with either the fixed amplitude or the regular $N$-body simulations. We will quantify the difference between these two choices below.
We first estimate the observational covariance matrix, $C_{\rm obs}$, by rescaling the covariance matrix from the regular simulations based on the expected survey volume. Consider an effective volume of 20 ($h^{-1}$Gpc)$^3$, roughly corresponding to that of the DESI and Euclid surveys. The covariance matrix including a pair of fixed-amplitude simulations can be computed by 
\begin{equation}
    C = C_{\rm SVM} + \frac{C_{1}}{V_{\rm EFFS}},
\end{equation}
where $C_1$ is the covariance matrix of a single regular 1($h^{-1}$Gpc)$^3$ box, $V_{\rm EFFS}$ is the effective survey volume (20 ($h^{-1}$Gpc)$^3$ in our study), and $C_{SVM}$ is the covariance matrix of the suppressed variance method (paired fixed amplitude simulation).
Following Equation \ref{eq:Hamachi}, we compute the variance, ${\rm Var}_{\rm SVM}$.

Let us now answer the question: What is the size of the required standard simulation, that yields the equivalent variance of a pair of simulations with the SVM? 
Given a normal simulation with volume V= ($h^{-1}$Gpc)$^3$, the total covariance matrix is given by
\begin{equation}
    C_V = \frac{C_{1}}{V} + \frac{C_{1}}{V_{\rm EFFS}}.
\end{equation}
We now compute the variance ${\rm Var}_{\rm V}$ based on \ref{eq:Hamachi}. By solving ${\rm Var}_{\rm SVM}$ = ${\rm Var}_{\rm V}$, we obtain the equivalent volume ($V$) that our paired fixed amplitude simulations are representing. This is shown in Fig. \ref{fig:improve_100par}; the equivalent volume vs. scale ranges used in the power spectrum and correlation function analysis are shown. Here the maximum separation was fixed and the minimum separation was varied in the correlation function analysis; while the minimum $k$ was fixed and the maximum $k$ was varied in the power spectrum analysis.

We find that a pair of 1 ($h^{-1}$Gpc)$^3$ boxes can potentially correspond to  effective volumes of up to 100 ($h^{-1}$Gpc)$^3$ considering  halos with lower masses. We also find that the equivalent volume is sensitive to the power spectrum, but not to the correlation function analysis.
One might obtain very large effective volumes by ignoring the covariance matrix from observations, artificially driven by the uncertainty at large scales (e.g. small $k$). Thus, this additional covariance matrix needs to be taken into account,  as we do in our analysis.

Interestingly, the naive correspondence between $k\sim 0.35\,h$Mpc$^{-1}$ and $r\sim 20 \,h^{-1}$Mpc using $k=2\pi/L$, yields completely different effective volumes: roughly 10 and 100  $(h^{-1}{\rm Gpc})^3$, respectively (see lower panels in Fig. \ref{fig:improve_100par}), thus emphasizing the difference in Fourier and configuration space analyses when a limited  range in $k$ or $r$ is used.

In contrast to configuration space, Fourier space is more sensitive to large scales (low $k$s), which are already linear (e.g. $k\sim0.2\,h$Mpc$^{-1}$).
Although fixing the amplitude of the power spectrum is crucial in reducing variance, as we showed in detail in Section \ref{sec:fastpm_results}, it does not remove variance induced by nonlinear gravitational mode-coupling.
This is very apparent in Fourier space analysis.
We can find analogous examples in the literature comparing the two-point statistics in Fourier and configuration space, such as 1) aliasing introduced by the gridding process of a set of point masses onto a mesh \citep[][]{1988csup.book.....H}, in which a clouds-in-cells mass assignment scheme applied on dark matter particles in a cosmological simulation with cell resolutions of a few Mpc scales underestimates the true power spectrum down to $k\sim0.2\,h$Mpc$^{-1}$ \citep{2005ApJ...620..559J}; 2) or in the clustering analysis of galaxies in redshift space, in which the virial motions (a.k.a. fingers-of-god, \citealt{1972MNRAS.156P...1J}) are present only below a certain scale (of $\sim20$ Mpc), but are visible down to $k\sim0.1\,h$Mpc$^{-1}$ in Fourier space.
This also indicates that pairing simulations with opposite phases and fixed amplitudes is not very effective in suppressing the variance at small scales, as we already saw in the three-point statistics analysis (see Section \ref{sec:fastpm_results}) and further improvements should be investigated.

We conclude from this analysis, that our two pairs of high resolution $N$-body simulations with the SVM have an effective volume larger than 7 times that of the DESI or Euclid effective survey volumes when the analysis is performed in configuration space. We are currently preparing larger sets of SVM $N$-body simulations to ensure that this accuracy is also achieved in Fourier space.

\section{Summary and conclusions}
\label{sec:conclusion}
In this work we have presented the UNIT $N$-body cosmological simulation project. 
We present four simulations (two pairs) along with this paper. The box size is $1 h^{-1} \rm{Gpc}$ and the number of particles of each box is $4096^3$, resulting in a particle mass of $\sim 1.2\times10^9$ $h^{-1}$M$_{\odot}$.
We have made their corresponding data products publicly available through the website \url{http://www.unitsims.org}, including
\begin{enumerate}
\item dark matter particles
\item density fields
\item halo catalogs
\item dark matter clustering statistics
\item halo clustering statistics (real and redshift space)
\item void clustering statistics
\end{enumerate}

We show that the effective volume of our simulation suite is equivalent to 150 ($h^{-1}$ Mpc)$^3$ (7 times of the effective survey volume of DESI or Euclid), using a mass resolution of $\sim 1.2\times10^9\,h^{-1}$M$_{\odot}$, enough to resolve the host halos of the galaxy sample observed by DESI (ELGs) or Euclid (H$\alpha$ galaxies).

Our work relies on the suppressed variance method (SVM) approach recently introduced by \cite{Angulo:2016hjd}.  In order to demonstrate the practicality of the SVM for large-scale structure analyses, we investigate a number of issues including potential biases introduced by the method, and characterize the improvement in the theoretical uncertainty and effective volume in a number of different regimes.

We have performed a large number (800) of accurate particle mesh simulations using the FastPM code, and have demonstrated that
\begin{itemize}
    \item no significant biases are introduced that would affect BAO or RSD analysis;
    \item the error in two-point statistics in configuration space is significantly reduced;
    \item the error in two-point statistics in Fourier space in moderately reduced; and
    \item no significant improvement is found for the three-point statistics on scales relevant to BAO and RSD analysis.
\end{itemize}

We also performed an analysis including redshift-space distortions, and three-dimensional halo distributions beyond the halo mass function.
We found that the improvements in galaxy bispectrum and void auto-correlation function using SVM are small. However, the improvement in the void--galaxy cross-correlation is significant;  this indicates that the fixed-amplitude method should also be useful for void studies.

We introduced a parameter for quantifying the improvement of the suppressed variance methods, and show that these simulations are equivalent to a typical simulation with volume of 100 ($h^{-1}$ Gpc)$^3$. The exact number depends on the analysis method considered (e.g., power spectrum or correlation function analysis), redshift, scale range, and the galaxy sample used.

With current state-of-the-art techniques we found that for a galaxy survey with effective volume of 20 ($h^{-1}$ Gpc)$^3$ at $z=1$, the reduction in variance resulting from the SVM is about a factor of 40 using two-point correlation function analysis. This means that our two pairs of simulations with full $N$-body calculations with volumes of (1$h^{-1}$ Gpc)$^3$ and 4096$^3$ particles lead to the same variance as $\sim$150 of such simulations. 
This provides optimal reference clustering measurements to validate theoretical models in configuration space. 
The covariance matrices for the clustering measurements using SVM can be estimated based on the approximate methods presented in this paper, since they are very different from the typical Gaussian statistics (see \citealt{Angulo:2016hjd}).
This motivates future work to 
compute larger sets of $N$-body simulations using SVM.  We are pursuing this, along with further analyses to investigate mode-coupling effects from larger scales and ways of correcting them.

In the spirit of sharing scientific results with the community, we have made 
the full $N$-body simulations in addition to the FastPM products produced in this work publicly available through the website \url{http://www.unitsims.org}.  We hope that these data products will enable a number of studies to further unveil the nature of dark energy and structure formation with galaxy surveys.


\section*{Acknowledgements}
We thank Raul E. Angulo, Elena Massara, and Andrew Pontzen for useful discussions, and Volker Springel for providing the L-Gadget code.
We acknowledge PRACE for awarding us access to  the MareNostrum Supercomputer  hosted by the  Barcelona Supercomputing Center, Spain,  under project grant  number 2016163937.
GY and AK  would like to thank MINECO/FEDER (Spain) for partial financial support under project grant AYA2015-63810P. AK further thanks the Spanish Red Consolider MultiDark FPA2017-90566-REDC.
FSK thanks support from the grants RYC2015-18693, SEV-2015-0548, and AYA2017-89891-P. CZ thanks support under NSFC Grant No. 11673025. SA is supported by the European Research Council through the COSFORM Research Grant (\#670193). This research received support from the U.S. Department of Energy under contract number DE-AC02-76SF00515,  Additional computing support was provided by SLAC and by National Energy Research Scientific Computing Center (NERSC), a U.S. Department of Energy Office of Science User Facility operated under Contract No. DE-AC02-05CH11231.




\bibliographystyle{mnras}
\bibliography{unitsim}





\bsp	
\label{lastpage}
\end{document}